\documentstyle[preprint,prd,aps]{revtex}
\begin{document}
\draft

\title{
One-loop background calculations in the general field theory.}

\author{Petr I.Pronin
\thanks{E-mail:$pronin@theor.phys.msu.su$}
and Konstantin V.Stepanyantz
\thanks{E-mail:$stepan@theor.phys.msu.su$}}

\address{
Moscow State University, Physical Faculty,\\
Department of Theoretical Physics.\\
$117234$, Moscow, Russian Federation}

\maketitle

\begin{abstract}
We present master formulas for the divergent part of the one-loop
effective action for a minimal operator of any order in the
4-dimensional curved space and for an arbitrary nonminimal operator
in the flat space.
\end{abstract}


\narrowtext

\unitlength=1pt
\sloppy


\section{Introduction.}

Progress in the quantum field theory and quantum gravity in particular
depends much on the development of methods for the calculation of the
effective action. For a lot of problems the analyses can be confined to the
one-loop approximation. In this case the effective action can be
expressed as a functional trace of an elliptic operator logarithm
\cite{jackiw}. The calculation of its divergent part in the dimensional
regularization is then reduced to finding $a_2$ coefficient of the
heat kernel expansion \cite{minak,seeley,gilkey}. It can be made in the
frames of some approaches, for example, by the proper-time method proposed by
Schwinger \cite{schwinger} and extended to the curved space-time by DeWitt
\cite{dewitt}. It allows to consider the simplest minimal second order
operator

\begin{eqnarray}
D_{1\ i}{}^j = \nabla^\mu \nabla_\mu \delta_i{}^j
+ S^\mu{}_i{}^j \nabla_\mu + W_i{}^j.
\end{eqnarray}

On the other hand, the effective action is the sum of one-particle
irreducible Feynman graphs. Explicitly covariant perturbation theory
can be constructed if only the propagator depends on the background metric.
Then the calculation of divergent graphs can be made by the generalized
Schwinger-DeWitt technick presented in \cite{barv}. Using this approach
Barvinsky and Vilkovisky found the effective action for the theories with
the operators

\begin{eqnarray}
&&D_{2\ i}{}^j = \left(\nabla^\mu \nabla_\mu\right)^2 \delta_i{}^j
+ S^{\mu\nu\alpha}{}_i{}^j \nabla_\mu \nabla_\nu \nabla_\alpha
+ W^{\mu\nu}{}_i{}^j \nabla_\mu \nabla_\nu\nonumber\\
&&\vphantom{\frac{1}{2}}
+ N^{\mu}{}_i{}^j \nabla_\mu + M_i{}^j, \nonumber\\
&&\vphantom{\frac{1}{2}}
D_{3\ \alpha}{}^\beta =
\nabla^\mu \nabla_\mu \delta_{\alpha}{}^{\beta}
- \lambda\ \nabla_\alpha \nabla^\beta + P_{\alpha}{}^\beta.
\end{eqnarray}

The method has been extended to multi-loop orders \cite{abbott,loop2},
non-Riemannian manifolds \cite{obukhov,barth,cogzer,gusynin,gusynin1}
and theories with fermions \cite{leerim,fermiloop2}. Some papers were
devoted to varies changes in the technick of calculations
\cite{cogzer,gusynin,gusynin1}.

Recently the approach proposed in \cite{barv} was applied for
the calculation of non-local contributions to the effective action
on the Riemannian manifold \cite{nonlocal}.

In this paper we construct the explicit expression for the divergent part
of the one-loop effective action without any restrictions to the form and
the order of the operator $D$. (We consider only theories in the
dimension 4). For this purpose we use the generalized t'Hooft-Veltman
approach \cite{thooft}.

Our paper is organized as follows.

In Sec. 2 we briefly describe the calculation of the one-loop
effective action by gaussian integration and introduce some notations.

In Sec. 3 we remind t'Hooft-Veltman diagramic technick and derive Feynman
rules for arbitrary minimal and nonminimal operators. We then found
all divergent graphs in the flat space and calculate their degree of
divergence.

In Sec. 4 we consider a theory with an arbitrary minimal operator
in the curved space-time. The main result here is an explicit expression
for the one-loop contribution to the divergent part of the effective action.

Section 5 is devoted to the derivation of a master formula for an arbitrary
nonminimal operator on the flat background.

In Sec. 6 we consider some particular cases and show the agreement
of our formulas with the earlier known results. We find also the
effective action for theories with sixth and eighth order minimal
operators.

In Sec. 7 we give a brief discussion and summary of our results.

In Appendix A we describe a method to calculate the divergent part
of Feynman integrals.

In Appendix B we derive some rules for handling the integration
over angles in the case of a minimal operator.


\section{Background field method in the one-loop approximation}

Let us consider the generating functional for connected Green's
functions

\begin{eqnarray}
exp\left(\frac{i}{\hbar} W[J]\right) = \int {\it D}\varphi\
exp\left(\frac{i}{\hbar}(S[\varphi] + J\varphi)\right).
\end{eqnarray}

\noindent
(In this section $\hbar \ne 1$.)

Its loop expansion may be obtained by the method of saddle-point
integration. The saddle point $\varphi_0[J]$ satisfy

\begin{eqnarray}
\left.\frac{\delta S}{\delta \varphi}\right|_{\varphi=\varphi_0} = -\ J.
\end{eqnarray}

Expanding $S$ about $\varphi_0$, we found that

\begin{eqnarray}\label{WJ}\label{wloop}
W[J] = S[\varphi_0] + J\varphi_0
+ \frac{i}{2}\ \hbar\ tr\ ln\ \frac{\delta^2 S}{\delta \varphi_0^2}
+ O(\hbar^2).
\end{eqnarray}

The effective action $\Gamma[\varphi]$ is defined by making Legender
transformation

\begin{eqnarray}\label{effectiveactiondef}
\Gamma[\varphi] = \left.W[J] - J\varphi\right|_{J=J[\varphi]}.
\end{eqnarray}

\noindent
Substituting (\ref{wloop}) into (\ref{effectiveactiondef})
we obtain the loop expansion of the effective action to one-loop order
\cite{jackiw}

\begin{eqnarray}
\Gamma[\varphi]
=S[\varphi]
+ \frac{i}{2}\ \hbar\ tr\ ln\ \frac{\delta^2 S}{\delta \varphi^2}
+ O(\hbar^2).
\end{eqnarray}

\noindent
If $\varphi$ is a complex field of an arbitrary tensor structure,
the effective action takes the form

\begin{eqnarray}
\Gamma^{(1)} = i\ \hbar\ tr\ ln
\frac{\delta^2 S}{\delta \varphi^{*i}\delta \varphi_j},
\end{eqnarray}

\noindent
where latin letters denote the whole set of $\varphi$ indexes.

\begin{eqnarray}
D_i{}^j \equiv
\frac{\delta^2 S}{\delta \varphi^{*i}\delta \varphi_j}
\end{eqnarray}

\noindent
is a differential operator depending on the background field $\varphi$.
Its most general form is

\begin{eqnarray}\label{nonminimal}
&&
\vphantom{\frac{1}{2}}
D_{i}{}^{j} =
K^{\mu_1\mu_2\ldots \mu_{L}}{}_{i}{}^{j}
\ \nabla_{\mu_1} \nabla_{\mu_2}\ldots
\nabla_{\mu_{L}}
\nonumber\\
&&
\vphantom{\frac{1}{2}}
+\ S^{\mu_1\mu_2\ldots \mu_{L-1}}{}_{i}{}^{j}
\ \nabla_{\mu_1} \nabla_{\mu_2}\ldots
\nabla_{\mu_{L-1}}
\nonumber\\
&&
\vphantom{\frac{1}{2}}
+\ W^{\mu_1\mu_2\ldots \mu_{L-2}}{}_{i}{}^{j}
\ \nabla_{\mu_1} \nabla_{\mu_2}\ldots
\nabla_{\mu_{L-2}}
\nonumber\\
&&
\vphantom{\frac{1}{2}}
+\ N^{\mu_1\mu_2\ldots \mu_{L-3}}{}_{i}{}^{j}
\ \nabla_{\mu_1} \nabla_{\mu_2}\ldots
\nabla_{\mu_{L-3}}
\nonumber\\
&&
\vphantom{\frac{1}{2}}
+\ M^{\mu_1\mu_2\ldots \mu_{L-4}}{}_{i}{}^{j}
\ \nabla_{\mu_1} \nabla_{\mu_2}\ldots
\nabla_{\mu_{L-4}} +
\ldots,
\end{eqnarray}

\noindent
where $\nabla_\mu$ is a covariant derivative:

\begin{eqnarray}
\nabla_\alpha T^\beta{}_i{}^j = \partial_\alpha T^\beta{}_i{}^j +
\Gamma_{\alpha\gamma}^\beta&& T^\gamma{}_i{}^j \nonumber\\
\vphantom{\frac{1}{2}}
+ \omega_{\alpha i}{}^k
&&T^\beta{}_k{}^j - T^\beta{}_i{}^k \omega_{\alpha k}{}^j,\nonumber\\
\vphantom{\frac{1}{2}}
\nabla_\mu \Phi_i = \partial_\mu \Phi_i + \omega_{\mu i}{}^j \Phi_j. \ &&
\end{eqnarray}

Here $\Gamma_{\mu\nu}^\alpha$ is the Cristoffel symbol

\begin{equation}
\Gamma_{\mu\nu}^\alpha = \frac{1}{2} g^{\alpha\beta} (\partial_\mu
g_{\nu\beta} + \partial_\nu g_{\mu\beta} - \partial_\beta g_{\mu\nu})
\end{equation}

\noindent
and $\omega_{\mu i}{}^j$ is a connection in the principle bundle.

Commuting covariant derivatives we can always make $K, S, W, N, M, \ldots$
symmetric in the greek indexes. This condition is very convenient for
the calculations, so we will assume it to be satisfied.

The operator is called minimal if $L = 2l$ and

\begin{eqnarray}
K^{\mu\nu\ldots\alpha}{}_i{}^j =
K_0{}^{\mu\nu\ldots\alpha}\delta_i{}^j,
\end{eqnarray}

\noindent
where $K_0$ is a totally symmetric tensor, built by $g_{\mu\nu}$:

\begin{equation}\label{k0}
K_0{}^{\mu\nu\alpha\beta \dots } = \frac{1}{(2l-1)!!}
(g_{\mu\nu} g_{\alpha\beta} \ldots  +
g_{\mu\alpha} g_{\nu\beta} \ldots  + \ldots ).
\end{equation}

\noindent
(Here the sum is over all possible index replacements). If an
operator can not be reduced to this form, we will call it nonminimal.

Commuting covariant derivatives we can rewrite a minimal operator in
another form:

\begin{eqnarray}\label{minimal1}
&&D_{i}{}^{j}\ =\ \delta_{i}{}^{j}\ \Box^l +
\ S^{\mu_1\mu_2\ldots \mu_{2l-1}}{}_{i}{}^{j}
\ \nabla_{\mu_1} \nabla_{\mu_2}\ldots
\nabla_{\mu_{2l-1}}
\nonumber\\
&&\vphantom{\frac{1}{2}}
+\ W^{\mu_1\mu_2\ldots \mu_{2l-2}}{}_{i}{}^{j}
\nabla_{\mu_1} \nabla_{\mu_2}\ldots
\nabla_{\mu_{2l-2}}
\nonumber\\
&&\vphantom{\frac{1}{2}}
+\ N^{\mu_1\mu_2\ldots \mu_{2l-3}}{}_{i}{}^{j}
\nabla_{\mu_1} \nabla_{\mu_2}\ldots
\nabla_{\mu_{2l-3}}
\nonumber\\
&&\vphantom{\frac{1}{2}}
+\ M^{\mu_1\mu_2\ldots \mu_{2l-4}}{}_{i}{}^{j}
\nabla_{\mu_1} \nabla_{\mu_2}\ldots
\nabla_{\mu_{2l-4}} +
\ldots,
\end{eqnarray}

\noindent
where $\Box \equiv \nabla_{\mu} \nabla^{\mu}$.


\section{Diagramic approach in the background field method}

We will calculate the divergent part of the one-loop
effective action

\begin{eqnarray}\label{effectiveaction}
\Gamma^{(1)} = \frac{i}{2}\ tr\ ln\ D_i{}^j
\end{eqnarray}

\noindent
for both arbitrary minimal and arbitrary nonminimal operators
by the diagramic technick.

In order to present (\ref{effectiveaction}) as a sum of one-loop diagrams
we separate a differential operator into a term with the largest
number of derivatives plus a perturbation term:

\begin{eqnarray}
D_i{}^j = \partial^{2l}\delta_i{}^j + V_i{}^j \quad
\end{eqnarray}

(minimal operator)

\begin{eqnarray}
D_i{}^j = K^{\mu\nu\ldots\alpha}{}_i{}^j
\ \partial\mu \partial_\nu \ldots\partial_\alpha + V_i{}^j
\end{eqnarray}

(nonminimal operator)

\noindent
where

\begin{eqnarray}\label{purt}
&&\vphantom{\frac{1}{2}}
V = S^{\mu_1\ldots\mu_{L-1}} \partial_{\mu_1}\ldots\partial_{\mu_{L-1}}
\nonumber\\
&&\vphantom{\frac{1}{2}}
+\ W^{\mu_1\ldots\mu_{L-2}} \partial_{\mu_1}\ldots\partial_{\mu_{L-2}}
\nonumber\\
&&\vphantom{\frac{1}{2}}
+\ N^{\mu_1\ldots\mu_{L-3}} \partial_{\mu_1}\ldots\partial_{\mu_{L-3}}
\nonumber\\
&&\vphantom{\frac{1}{2}}
+\ M^{\mu_1\ldots\mu_{L-4}} \partial_{\mu_1}\ldots\partial_{\mu_{L-4}}
+ \ldots + O(h,\omega).
\end{eqnarray}

\noindent
Terms $O(h,\omega)$ can be found by a series expansion of the operator
in powers of $h_{\mu\nu} = g_{\mu\nu} - \eta_{\mu\nu}$ and
$\omega_\mu{}_i{}^j$.

Then the one-loop effective action  (\ref{effectiveaction}), say, for a
minimal operator can be written as

\begin{eqnarray}\label{efac}
\Gamma^{(1)}&& = \frac{i}{2}\ tr\ ln\ D_i{}^j =
\frac{i}{2}\ tr\ ln\ \left(\partial^{2l} + V \right)
\nonumber\\
&&
=\frac{i}{2}\ tr\ ln\ \partial^{2l}\
+\ \frac{i}{2}\ tr\ ln\ \left(1 + \frac{1}{\partial^{2l}} V\right).
\end{eqnarray}

First term in (\ref{efac}) is an infinite numeric constant and can be
omitted. Expanding logarithm in the second term in powers of $V$ we
obtain

\begin{eqnarray}\label{purttheor}
\Gamma^{(1)} =
\frac{i}{2}\ tr\ ln\ &&\left(1 + \frac{1}{\partial^{2l}} V\right)\nonumber\\
&&= \frac{i}{2}\ tr \sum_{k=1}^\infty \frac{1}{k}
\left(-\ \frac{1}{\partial^{2l}} V\right)^k.
\end{eqnarray}

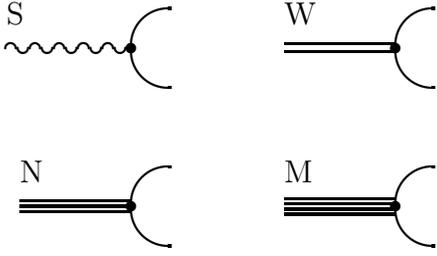
\begin{figure}
\begin{picture}(160,120)(0,0)
\thicklines

\put(90,90){\oval(30,30)[l]}
\put(75,90){\circle*{4.2}}
\put(29.5,90)
{\thicklines
\multiput(0,0)(9.2,0){5}%
{\put(0,-0.05){\oval(4,4)[t]}
\put(4.6,0.05){\oval(4,4)[b]}}}
\put(28,99){S}

\put(190,90){\oval(30,30)[l]}
\put(175,90){\circle*{4.2}}
\put(133,91.5){\line (1,0) {42}}
\put(133,88.5){\line (1,0) {42}}
\put(133,99){W}

\put(90,30){\oval(30,30)[l]}
\put(75,30){\circle*{4.2}}
\put(30,30){
\put(3,2.1){\line (1,0) {42}}
\put(3,0){\line (1,0) {42}}
\put(3,-2.1){\line (1,0) {42}}}
\put(33,39){N}

\put(190,30){\oval(30,30)[l]}
\put(175,30){\circle*{4.2}}
\put(130,30){
\put(3,3){\line (1,0) {42}}
\put(3,-3){\line (1,0) {42}}
\put(3,1){\line (1,0) {42}}
\put(3,-1){\line (1,0) {42}}}
\put(133,39){M}

\end{picture}

\caption{Vertexes, that are encountered in the flat space
divergent graphs.}\label{vertexes}
\end{figure}

(\ref{purttheor}) can be presented as a sum of one-loop diagrams
with $k$ vertexes. In the momentum space the propagator has the form
$\delta_{i}^{j} /k^{2l}$. Vertexes in the flat space are presented
at the Fig.~\ref{vertexes}. Their expressions are rather evident, for
example, the vertex with $S$-external line can be written as

\begin{eqnarray}
S^{\mu_1\mu_2\ldots\mu_{L-1}} k_{\mu_1} k_{\mu_2} \ldots k_{\mu_{L-1}}
\equiv (Sk).
\end{eqnarray}

Similar notations we will use for other expressions, for example,

\begin{eqnarray}
&&(W(k+p))^\alpha \equiv
W^{\mu\nu\ldots \beta\alpha} (k+p)_\mu (k+p)_\nu\ldots (k+p)_\beta,
\nonumber\\
&&(Sk) ^{\alpha\beta}\ \equiv
S^{\mu\nu\ldots \gamma\alpha\beta} k_\mu k_\nu\ldots k_\gamma.
\end{eqnarray}

Numerical factors for the Feynman graphs can be easily found by
(\ref{purttheor}).

The number of diagrams in (\ref{purttheor}) is infinite, but
most of them are convergent. Really, it is easy to see that
the degree of divergence of a one-loop graph with $s$ $S$-vertexes,
$w$ $W$-vertexes, $n$ $N$-vertexes, $m$ $M$-vertexes and so on
($k = s+w+n+m+\ldots$) in the flat space ($h_{\mu\nu}=0$,
$\omega_\mu{}_i{}^j=0$) is

\begin{eqnarray}\label{degree}
I = 4 - s - 2 w - 3 n -4 m - \ldots.
\end{eqnarray}

\noindent
Therefore, there are only a finite number of the divergent diagrams.
They are presented at the Fig.~\ref{flatdiagrams}. (We excluded divergent
graphs, that give zero contribution to the effective action.)

(\ref{degree}) is valid also for a nonminimal operator. In this case
the propagator takes the form

\begin{eqnarray}
\frac{1}{K^{\mu_1\ldots\mu_L}{}_i{}^j \partial_{\mu_1} \ldots
\partial_{\mu_L}}
\end{eqnarray}

\noindent
in the coordinate space, or $(Kk)^{-1}{}_i{}^j$ in the momentum space,
where

\begin{eqnarray}
&&(Kk)_i{}^j\ \equiv K^{\mu\nu\ldots \alpha}{}_i{}^j
\ k_\mu k_\nu\ldots k_\alpha,\\
&&
\vphantom{\frac{1}{2}}
(Kk)^{-1}{}_i{}^m\ (Kk)_m{}^j = \delta_i{}^j.\nonumber
\end{eqnarray}

\noindent
We will assume that $(Kk)^{-1}{}_{i}{}^{j}$ exists. It can be made
by fixing gauge invariances.


\section{Effective action for a minimal operator}

\subsection{Flat space}

Now we should calculate diagrams presented at the Fig.~\ref{flatdiagrams}.
We will do it using dimensional regularization. So, in order to find the
divergent part of an integral

\begin{equation}
\int d^{d}k\ f(k,p)
\end{equation}

\noindent
it is necessary to expand the function $f$ into series, retain only
logarithmically divergent terms and perform the integration in remaining
integrals

\[\int d^{d}k\ \frac{1}{k^{M+4}}
k_{\mu_{1}} k_{\mu_{2}} \ldots  k_{\mu_{M}} \]

\noindent
according to the following equations

\begin{eqnarray}\label{angleintegration}
&&\int d^{d}k\ \frac{1}{k^{2m+5}}
k_{\mu_{1}} k_{\mu_{2}} \ldots  k_{\mu_{2m+1}} = 0,\nonumber\\
&&\int d^{d}k\ \frac{1}{k^{2m+4}}
k_{\mu_{1}} k_{\mu_{2}} \ldots  k_{\mu_{2m}} \nonumber\\
&&= -\ \frac{2 i\pi^2}{(d-4)} <n_{\mu_1} n_{\mu_2} \ldots n_{\mu_{2m}}>,
\nonumber
\end{eqnarray}

\noindent
where

\begin{eqnarray}\label{angle}
< n_{\mu_{1}} n_{\mu_{2}} \ldots n_{\mu_{2m}}&&>\ \equiv
\frac{1}{2^{m} (m+1)!}
\nonumber\\
\times
\vphantom{\frac{1}{2}}
\left(g_{\mu_{1}\mu_{2}} g_{\mu_{3}\mu_{4}}
\right. \ldots &&g_{\mu_{2m-1}\mu_{2m}}
\nonumber\\
+&&
\vphantom{\frac{1}{2}}
\left.
g_{\mu_{1}\mu_{3}} g_{\mu_{2}\mu_{4}}\ldots g_{\mu_{2m-1}\mu_{2m}} + \ldots
\right)
\end{eqnarray}

\noindent
is a result of the integration over angles (for more details see appendix
\ref{integ}).

(The sum in the equation (\ref{angle}) is over all possible index
replacements)

Let us consider first logarithmically divergent graphs (2a)-(2e).
In order to find the divergent part of the diagram in this case
we should retain only terms without external momentums and perform
the remaining integration. As an example let us calculate a
diagram (2d).

\begin{equation}
(2d) = \frac{i}{2(2\pi)^4}
\ tr \int d^dk\ \frac{(Sk)\ (N(k-p))}{k^{2l} (k-p)^{2l}}.
\end{equation}

\noindent
According to the previous discussion we can easily conclude that

\begin{eqnarray}\label{diag1}
(2d)_\infty
= \frac{i}{2(2\pi)^4}\ &&\int d^dk\ \frac{1}{k^4}\ tr <\hat S\ \hat N>
\nonumber\\
&&\ = \frac{1}{16\pi^2(d-4)}\ tr <\hat S\ \hat N>,
\end{eqnarray}

\noindent
where

\begin{eqnarray}\label{notations}
&&\hat S \ \equiv (Kn)^{-1} (Sn),
\nonumber\\
&&
\vphantom{\frac{1}{2}}
\hat N\ \equiv (Kn)^{-1} (Nn),
\end{eqnarray}

\noindent
and  $n_\mu = k_\mu/\sqrt{k^\alpha k_\alpha}$ is a unit vector.
In particular, for a minimal operator

\begin{eqnarray}
&&\hat S \ = (Sn)\ = S^{\mu\nu\ldots \alpha} n_\mu n_\nu\ldots n_\alpha
,\\
&&
\vphantom{\frac{1}{2}}
\hat N\ = (Nn) \ =
N^{\mu\nu\ldots \alpha} n_\mu n_\nu\ldots n_\alpha.\nonumber
\end{eqnarray}

\noindent
In a similar fashion we have

\begin{eqnarray}\label{ln}
&&(2a)_\infty = \frac{1}{64\pi^2(d-4)}\ tr <\hat S^4>,
\nonumber\\
&&(2b)_\infty = -\ \frac{1}{16\pi^2(d-4)}\ tr <\hat W\ \hat S^2>,
\nonumber\\
&&(2c)_\infty = \frac{1}{32\pi^2(d-4)}\ tr <\hat W^2>,
\nonumber\\
&&(2e)_\infty = -\ \frac{1}{16\pi^2(d-4)}\ tr <\hat M>.
\end{eqnarray}

The calculation of linearly divergent graphs is a bit more difficult.
For example, in order to find the divergent part of the diagram

\begin{equation}
(2g) = \frac{i}{2(2\pi)^4}
\ tr \int d^dk\ \frac{(Sk)\ (W(k-p))}{k^{2l} (k-p)^{2l}}
\end{equation}

\noindent
we should retain only terms linear in external momentum $p$. Using the
rule (\ref{eqa}) we obtain

\widetext

\begin{eqnarray}
\frac{1}{16\pi^2(d-4)}
\ tr <2l\ (pn)\ &&\hat S\ \hat W - (2l-2)\ p_\mu \hat W^\mu \hat S>
\nonumber\\
&&
= \frac{1}{32\pi^2(d-4)}
\ tr<(2l-1)\ p_\mu \hat S^\mu \hat W
- (2l-2)\ p_\mu \hat S\ \hat W^\mu>.
\end{eqnarray}

After a substitution $p_\mu \hat S \rightarrow -\ \partial_\mu \hat S$,
the result for this diagram takes the form

\begin{equation}
(2g)_\infty = \frac{1}{32\pi^2(d-4)}
\ tr <-\ (2l-1)\ \partial_\mu \hat S^\mu \hat W +
(2l-2)\ \partial_\mu \hat S\ \hat W^\mu >.
\end{equation}

\narrowtext

The second linearly divergent graph

\begin{equation}
(2h) = -\ \frac{i}{6(2\pi)^4}\ tr \int d^dk
\ \frac{
\stackrel{\displaystyle (Sk)}{\scriptscriptstyle (-p)\phantom{+}}
\stackrel{\displaystyle (S(k+q))}{\scriptscriptstyle (-q)
\phantom{+qqqqqqq}}
\stackrel{\displaystyle (S(k-p))}{\scriptscriptstyle (p+q)
\phantom{+qqqqqqq}}}
{k^{2l}
(k-p)^{2l} (k+q)^{2l}}
\end{equation}

\noindent
can be calculated in the same way. (Here indexes in the bottom point
the argument of $S$). The answer is

\begin{equation}
(2h)_\infty = \frac{(2l-1)}{48\pi^2(d-4)}
\ tr <\partial_\mu \hat S^\mu \hat S\ \hat S -
\partial_\mu \hat S\ \hat S\ \hat S^\mu>.
\end{equation}

So, we should consider only the rest quadratically divergent diagram.

\begin{equation}
(2f) = \frac{i}{4(2\pi)^4}
tr \int d^dk\ \frac{(Sk)(S(k-p))}{k^{2l}(k-p)^{2l}}.
\end{equation}

\noindent
Retaining logarithmically divergent terms we can easily find that

\widetext

\begin{eqnarray}\label{diag8}
(2f)_\infty =
\frac{1}{16\pi^2(d-4)}\ tr \int d^4 x
<\frac{l^2}{2(2l+1)}\ \partial_\mu \hat S\ \partial^\mu \hat S
+\frac{(2l-1)^2 l}{4(2l+1)}\ \partial_\mu \hat S^\mu \partial_\nu
\hat S^\nu&&
\nonumber\\
-\ \frac{(2l-1)(l^2-1)}{2(2l+1)}\ \partial_\mu &&\hat S^{\mu\nu}
\partial_\nu \hat S>.
\end{eqnarray}

Collecting the results (\ref{diag1})-(\ref{diag8}) we obtain
the divergent part of the one-loop effective action for the minimal
operator (\ref{minimal1}) in the flat space:

\begin{eqnarray}\label{flat}
&&\left( \Gamma^{(1)}_\infty \right)^{flat}=
\frac{1}{16\pi^2(d-4)}\ tr \int d^4 x\
<\frac{l^2}{2(2l+1)} \partial_\mu \hat S\ \partial^\mu \hat S
-\frac{(2l-1)(l^2-1)}{2(2l+1)}
\partial_\mu \hat S^{\mu\nu} \partial_\nu \hat S
\nonumber\\
&&
+\frac{(2l-1)^2 l}{4(2l+1)} \partial_\mu \hat S^\mu \partial_\nu \hat S^\nu
+\frac{(2l-1)}{3} \partial_\mu \hat S^\mu \hat S\ \hat S
-\frac{(2l-1)}{3} \partial_\mu \hat S\ \hat S\ \hat S^\mu
+(l-1)\ \partial_\mu \hat S\ \hat W^\mu
\\
&&
\vphantom{\frac{1}{2}}
- \frac{(2l-1)}{2} \partial_\mu \hat S^\mu \hat W
+\hat S\ \hat N + \frac{1}{2} \hat W^2 + \frac{1}{4} \hat S^4
- \hat W\ \hat S^2-\hat M>.\nonumber
\end{eqnarray}

\narrowtext


\subsection{Extension to the curved space}

In order to obtain the divergent part of a one-loop effective action in
the curved space we consider a minimal operator in the form
(\ref{minimal1}).

In this case we can not calculate all divergent graphs, because
their number is infinite. (The matter is that the degree of divergence
does not depend on the number of $h_{\mu\nu}$ vertexes and there are
infinite number of such vertexes too). Nevertheless if we note that the
answer should be invariant under the general coordinate transformations,
the result can be found by calculating only a finite number of graphs.
Really, we should replace derivatives in (\ref{flat}) by the covariant
ones and add expressions, containing curvature tensors
$R^\sigma{}_{\alpha\mu\nu}$ and $F_{\mu\nu}$. The most general form of
additional terms is

\widetext

\begin{eqnarray}\label{anzats}
&&\vphantom{\frac{1}{2}}
\frac{1}{16\pi^2(d-4)}\ tr \int d^4 x\ \sqrt{-g}
< a_1 R^2 + a_2 R_{\mu\nu} R^{\mu\nu}
+ a_3 \hat W^{\alpha\beta} R_{\alpha\beta}
+ a_4 \hat W R
+ a_5 \nabla_\mu \hat S^{\mu\alpha\beta} R_{\alpha\beta}
\nonumber\\
&&\vphantom{\frac{1}{2}}
+ a_6 \nabla_\mu \hat S^{\mu} R
+ a_7 \hat S^2 R
+ a_8 R_{\alpha\beta} \hat S^\alpha \hat S^\beta
+ a_9 R_{\alpha\beta} \hat S^{\alpha\beta} \hat S
+ a_{10} R_{\mu\nu\alpha\beta} \hat S^{\mu\alpha} \hat S^{\nu\beta}
+ a_{11} F_{\mu\nu} F^{\mu\nu}
\\
&&\vphantom{\frac{1}{2}}
+ a_{12} F_{\mu\nu} \hat S^\mu \hat S^\nu
+ a_{13} F_{\mu\nu} \nabla^\mu \hat S^\nu
>.\nonumber
\end{eqnarray}

\narrowtext

\noindent
where

\begin{eqnarray}
&&\vphantom{\frac{1}{2}}
R^{\alpha}{}_{\beta\mu\nu} =
\partial_\mu \Gamma_{\nu\beta}^{\alpha} -
\partial_\nu \Gamma_{\mu\beta}^{\alpha} +
\Gamma_{\mu\gamma}^{\alpha} \Gamma_{\nu\beta}^{\gamma} -
\Gamma_{\nu\gamma}^{\alpha} \Gamma_{\mu\beta}^{\gamma},\nonumber\\
&&\vphantom{\frac{1}{2}}
F_{\mu\nu i}{}^j =
\partial_\mu \omega_{\nu i}{}^j -
\partial_\nu \omega_{\mu i}{}^j +
\omega_{\mu i}{}^k \omega_{\nu k}{}^j -
\omega_{\nu i}{}^k \omega_{\mu k}{}^j,\nonumber\\
&&\vphantom{\frac{1}{2}}
R_{\mu\nu} = R^{\alpha}{}_{\mu\alpha\nu},
\qquad R = g^{\mu\nu} R_{\mu\nu}.
\end{eqnarray}

\noindent
(We take into account that the expression
$R_{\mu\nu\alpha\beta} R^{\mu\nu\alpha\beta} -
4 R_{\mu\nu} R^{\mu\nu} + R^2$
is a total derivative and should be omitted).

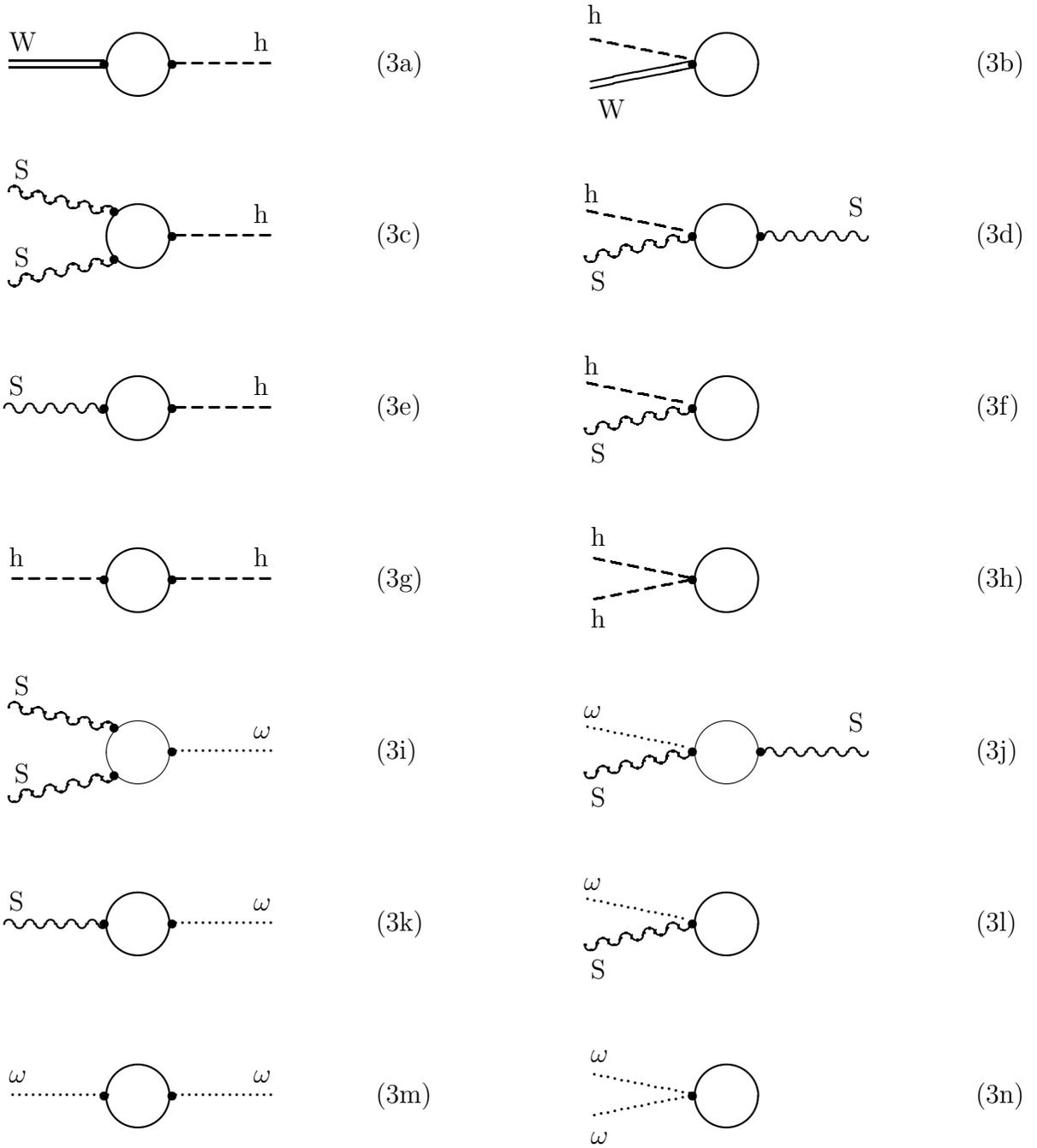
\begin{figure}

\begin{picture}(360,75)(50,0)
\thicklines

\put(110,30){\circle{29.9}}
\put(95,30){\circle*{4.2}}
\put(125,30){\circle*{4.2}}
\put(125,30){
\multiput(0,0)(7.5,0){6}%
{\multiput(0,0)(0.3,0){15}%
{.}}}
\put(53,31.5){\line (1,0) {42}}
\put(53,28.5){\line (1,0) {42}}
\put(53,36){W}
\put(161,36){h}
\put(215,27){(3a)}

\put(370,30){\circle{29.9}}
\put(355,30){\circle*{4.2}}
\put(308.5,40.5){
\multiput(0,0)(7.5,-1.5){6}%
{\multiput(0,0)(0.3,-0.06){15}%
{.}}}
\put(310,22.5){\line (5,1) {45}}
\put(310,19.5){\line (5,1) {45}}
\put(308.5,48){h}
\put(313,6){W}
\put(480,27){(3b)}
\end{picture}

\begin{picture}(360,75)(50,0)
\thicklines

\put(110,30){\circle{29.9}}
\put(99.5,19.5){\circle*{4.2}}
\put(99.5,40.5){\circle*{4.2}}
\put(125,30){\circle*{4.2}}
\put(55.4,49.5)
{\thicklines
\multiput(0,0)(10.3,-2.2){5}%
{\put(0,0){\oval(5,5)[t]}}
\multiput(5.3,-0.3)(10.3,-2.2){4}%
{\put(0,0){\oval(5,5)[b]}}
\multiput(1.8,-0.35)(10.3,-2.2){4}%
{.}}
\put(55.4,10.5)
{\thicklines
\multiput(0,0)(10.3,2.2){5}%
{\put(0,0){\oval(5,5)[b]}}
\multiput(5.3,0.3)(10.3,2.2){4}%
{\put(0,0){\oval(5,5)[t]}}
\multiput(1.6,0.1)(10.3,2.2){4}%
{.}}
\put(125,30){
\multiput(0,0)(7.5,0){6}%
{\multiput(0,0)(0.3,0){15}%
{.}}}
\put(55.4,16.5){S}
\put(55.4,55.5){S}
\put(161,36){h}
\put(215,27){(3c)}

\put(370,30){\circle{29.9}}
\put(355,30){\circle*{4.2}}
\put(385,30){\circle*{4.2}}
\put(310,21)
{\thicklines
\multiput(0,0)(10.3,2.2){5}%
{\put(0,0){\oval(5,5)[b]}}
\multiput(5.3,0.3)(10.3,2.2){4}%
{\put(0,0){\oval(5,5)[t]}}
\multiput(1.6,0.1)(10.3,2.2){4}%
{.}}
\put(307,40.5){
\multiput(0,0)(7.5,-1.5){6}%
{\multiput(0,0)(0.3,-0.06){15}%
{.}}}
\put(310,6){S}
\put(307,45){h}
\put(389,30)
{\thicklines
\multiput(0,0)(9.2,0){5}%
{\put(0,-0.05){\oval(4,4)[t]}
\put(4.6,0.05){\oval(4,4)[b]}}}
\put(424,39){S}
\put(480,27){(3d)}

\end{picture}

\begin{picture}(360,75)(50,0)
\thicklines

\put(110,30){\circle{29.9}}
\put(95,30){\circle*{4.2}}
\put(125,30){\circle*{4.2}}
\put(53,30)
{\thicklines
\multiput(0,0)(9.2,0){5}%
{\put(0,-0.05){\oval(4,4)[t]}
\put(4.6,0.05){\oval(4,4)[b]}}}
\put(125,30){
\multiput(0,0)(7.5,0){6}%
{\multiput(0,0)(0.3,0){15}%
{.}}}
\put(53,36){S}
\put(161,36){h}
\put(215,27){(3e)}

\put(370,30){\circle{29.9}}
\put(355,30){\circle*{4.2}}
\put(310,21)
{\thicklines
\multiput(0,0)(10.3,2.2){5}%
{\put(0,0){\oval(5,5)[b]}}
\multiput(5.3,0.3)(10.3,2.2){4}%
{\put(0,0){\oval(5,5)[t]}}
\multiput(1.6,0.1)(10.3,2.2){4}%
{.}}
\put(307,40.5){
\multiput(0,0)(7.5,-1.5){6}%
{\multiput(0,0)(0.3,-0.06){15}%
{.}}}
\put(310,6){S}
\put(307,45){h}
\put(480,27){(3f)}

\end{picture}

\begin{picture}(360,75)(50,0)
\thicklines

\put(110,30){\circle{29.9}}
\put(95,30){\circle*{4.2}}
\put(125,30){\circle*{4.2}}
\put(53,30){
\multiput(0,0)(7.5,0){6}%
{\multiput(0,0)(0.3,0){15}%
{.}}}
\put(125,30){
\multiput(0,0)(7.5,0){6}%
{\multiput(0,0)(0.3,0){15}%
{.}}}
\put(161,36){h}
\put(53,36){h}
\put(215,27){(3g)}

\put(370,30){\circle{29.9}}
\put(355,30){\circle*{4.2}}
\put(310,21){
\multiput(0,0)(7.5,1.5){6}%
{\multiput(0,0)(0.3,0.06){15}%
{.}}}
\put(310,39){
\multiput(0,0)(7.5,-1.5){6}%
{\multiput(0,0)(0.3,-0.06){15}%
{.}}}
\put(310,9){h}
\put(310,45){h}
\put(480,27){(3h)}

\end{picture}

\begin{picture}(360,75)(50,0)
\put(110,30){\circle{29.9}}
\put(99.5,19.5){\circle*{4.2}}
\put(99.5,40.5){\circle*{4.2}}
\put(125,30){\circle*{4.2}}
\put(55.4,49.5)
{\thicklines
\multiput(0,0)(10.3,-2.2){5}%
{\put(0,0){\oval(5,5)[t]}}
\multiput(5.3,-0.3)(10.3,-2.2){4}%
{\put(0,0){\oval(5,5)[b]}}
\multiput(1.8,-0.35)(10.3,-2.2){4}%
{.}}
\put(55.4,10.5)
{\thicklines
\multiput(0,0)(10.3,2.2){5}%
{\put(0,0){\oval(5,5)[b]}}
\multiput(5.3,0.3)(10.3,2.2){4}%
{\put(0,0){\oval(5,5)[t]}}
\multiput(1.6,0.1)(10.3,2.2){4}%
{.}}
\put(125,30){
\multiput(0,0)(3,0){15}%
{.}}
\put(55.4,16.5){S}
\put(55.4,55.5){S}
\put(161,36){$\omega$}
\put(215,27){(3i)}

\put(370,30){\circle{29.9}}
\put(355,30){\circle*{4.2}}
\put(385,30){\circle*{4.2}}
\put(310,21)
{\thicklines
\multiput(0,0)(10.3,2.2){5}%
{\put(0,0){\oval(5,5)[b]}}
\multiput(5.3,0.3)(10.3,2.2){4}%
{\put(0,0){\oval(5,5)[t]}}
\multiput(1.6,0.1)(10.3,2.2){4}%
{.}}
\put(307,40.5){
\multiput(0,0)(3,-0.6){15}%
{.}}
\put(310,6){S}
\put(307,45){$\omega$}
\put(389,30)
{\thicklines
\multiput(0,0)(9.2,0){5}%
{\put(0,-0.05){\oval(4,4)[t]}
\put(4.6,0.05){\oval(4,4)[b]}}}
\put(424,39){S}
\put(480,27){(3j)}

\end{picture}

\begin{picture}(360,75)(50,0)
\thicklines

\put(110,30){\circle{29.9}}
\put(95,30){\circle*{4.2}}
\put(125,30){\circle*{4.2}}
\put(53,30)
{\thicklines
\multiput(0,0)(9.2,0){5}%
{\put(0,-0.05){\oval(4,4)[t]}
\put(4.6,0.05){\oval(4,4)[b]}}}
\put(125,30){
\multiput(0,0)(3,0){15}%
{.}}
\put(53,36){S}
\put(161,36){$\omega$}
\put(215,27){(3k)}

\put(370,30){\circle{29.9}}
\put(355,30){\circle*{4.2}}
\put(310,21)
{\thicklines
\multiput(0,0)(10.3,2.2){5}%
{\put(0,0){\oval(5,5)[b]}}
\multiput(5.3,0.3)(10.3,2.2){4}%
{\put(0,0){\oval(5,5)[t]}}
\multiput(1.6,0.1)(10.3,2.2){4}%
{.}}
\put(307,40.5){
\multiput(0,0)(3,-0.6){15}%
{.}}
\put(310,6){S}
\put(307,45){$\omega$}
\put(480,27){(3l)}

\end{picture}

\begin{picture}(360,75)(50,0)
\thicklines

\put(110,30){\circle{29.9}}
\put(95,30){\circle*{4.2}}
\put(125,30){\circle*{4.2}}
\put(53,30){
\multiput(0,0)(3,0){15}%
{.}}
\put(125,30){
\multiput(0,0)(3,0){15}%
{.}}
\put(161,36){$\omega$}
\put(53,36){$\omega$}
\put(215,27){(3m)}

\put(370,30){\circle{29.9}}
\put(355,30){\circle*{4.2}}
\put(310,21){
\multiput(0,0)(3,0.6){15}%
{.}}
\put(310,39){
\multiput(0,0)(3,-0.6){15}%
{.}}
\put(310,9){$\omega$}
\put(310,45){$\omega$}
\put(480,27){(3n)}
\end{picture}

\caption{Graphs for the effective action calculation in the curved space.}
\label{curveddiagrams}
\end{figure}

Then the coefficients $a_1$ - $a_{13}$ can be found by calculating
the diagrams presented at the Fig.~\ref{curveddiagrams}. They conform
to the first nontrivial approximation in the counterterm expansion
in powers of weak fields $h_{\mu\nu}=g_{\mu\nu}-\eta_{\mu\nu}$ and
$\omega_{\mu i}{}^j$.

As an example we consider diagrams (3a) and (3b).

The vertex with $h_{\mu\nu}$ in (3a) should be found according to
(\ref{purt}) by series expansion of $\Box^l$ to the first order and
has the form

\widetext

\begin{eqnarray}
\sum_{m=0}^{l-1} k^{2m} (k-p)^{2l-2m-2}
\left(-h^{\mu\nu} k_\mu (k-p)_\nu + \frac{1}{2} h^\alpha{}_\alpha
p^\mu (k-p)_\mu \right).
\end{eqnarray}

Then the graph (3a) can be written as

\begin{eqnarray}\label{a3}
(3a) = \frac{i}{2(2\pi)^4}\ tr \int d^d k\ &&\frac{1}{k^{2l} (k-p)^{2l}}
\ (Wk)\nonumber\\
&&\times
\sum_{m=0}^{l-1} k^{2m} (k-p)^{2l-2m-2}
\left(-h^{\mu\nu} k_\mu (k-p)_\nu + \frac{1}{2} h^\alpha{}_\alpha
p^\mu (k-p)_\mu \right).
\end{eqnarray}

It is easy to see that the considered diagram is quadratically divergent.
So, retaining terms quadratic in external momentum $p$ we obtain the
divergent part

\begin{eqnarray}
\frac{1}{16\pi^2 (d-4)}\ &&tr <\hat W
\left(
l(l+1)\ (n^\gamma p_\gamma)
\left(h^{\mu\nu} n_\mu p_\nu +
\ \frac{1}{2} h^\alpha{}_\alpha (n^\beta p_\beta)\right)
\right.
-\nonumber\\
&&
\left.
-\ h^{\mu\nu} n_\mu n_\nu \left(
-\ \frac{1}{2} l(l+1)\ p^2
+\ \frac{2}{3} l(l+1)(l+2)\ (n^\alpha p_\alpha)^2\right)
-\ \frac{l}{2} h^\alpha{}_\alpha p^2
\right)>.
\end{eqnarray}

\noindent
Using rules formulated in the appendix \ref{diagangle}, it can be
written as

\begin{eqnarray}\label{w1}
&&(3a)_\infty = \frac{1}{16\pi^2(d-4)}
\ tr <
- \frac{1}{12} (2l-3)(2l-4)(2l-5)
\ \hat W^{\mu\nu\alpha\beta}
h_{\alpha\beta} p_\mu p_\nu
\nonumber\\
&&
+ \frac{l}{6}\hat W(-h^\alpha{}_\alpha p^2
+ h^{\mu\nu} p_\mu p_\nu)
- \frac{1}{12} (l-1)(2l-3)\ \hat W^{\mu\nu}
(- p^2 h_{\mu\nu} - p_\mu p_\nu h^\alpha{}_\alpha
+ 2 p_\mu p_\alpha h^\alpha{}_\nu)>.
\end{eqnarray}

Nevertheless, (\ref{w1}) can not be presented as a weak field limit
of a covariant expression. The matter is that the graphs (3a) and
(3b) can not be considered separately.

The vertex in the tadpole diagram (3b) can be found by series expansion of
$W^{\mu\nu\ldots\alpha}{}_i{}^j \nabla_\mu \nabla_\nu \ldots \nabla_\alpha$
in powers of $h_{\mu\nu}$ to the first order. Then, retaining only
nontrivial contributions, we have

\begin{eqnarray}\label{b3}
(3b) = \frac{i}{2(2\pi)^4}\ tr &&\int d^d k
\frac{1}{k^{2l}}
\ W{}^{\mu_1\mu_2 \ldots \mu_{2l-2}}
\nonumber\\
\times
&&\sum_{m=0}^{2l-4} (2l-3-m)
(k+p)_{\mu_1} \ldots (k+p)_{\mu_m}
\Gamma^{\alpha\ {\scriptscriptstyle (1)}}_{\mu_{m+1}\mu_{m+2}}
k{}_\alpha
k{}_{\mu_{m+3}} \ldots k_{\mu_{2l-2}},
\end{eqnarray}

\narrowtext

\noindent
where

\begin{equation}
\Gamma^{\alpha\ {\scriptscriptstyle (1)}}_{\beta\gamma} =
\frac{1}{2} (p_\beta h_\gamma{}^\alpha + p_\gamma h_\beta{}^\alpha
- p^\alpha h_{\beta\gamma})
\end{equation}

\noindent
is the Cristoffel symbol in the weak field limit.

Using (\ref{rule2}) after simple transformations the divergent part
of (3b) can be written as

\begin{eqnarray}\label{w2}
(3b)_\infty = \frac{(2l-3)(2l-4)(2l-5)}{192\pi^2(d-4)}&&
\nonumber\\
\times
\ tr< \hat W^{\mu\nu\alpha\beta} &&h_{\alpha\beta} p_\mu p_\nu>.
\end{eqnarray}

\noindent
The sum of (\ref{w1}) and (\ref{w2}) unlike each of the graphs
considered separately can be presented as a weak field approximation
of a covariant expression

\begin{eqnarray}
\frac{1}{16\pi^2(d-4)}
&&\ tr \int d^4 x\ \sqrt{-g}
<\frac{l}{6}\hat W R
\nonumber\\
&&- \frac{1}{6}(l-1)(2l-3)\ R_{\mu\nu} \hat W^{\mu\nu}>,
\end{eqnarray}

\noindent
that is the ultimate answer for the diagrams (3a) and (3b).

The other graphs are considered in the same way. After rather cumbersome
calculations we obtain the following formula for the divergent part of
the one-loop effective action for a minimal operator (\ref{minimal1}) in
the curved space:

\widetext

\begin{eqnarray}\label{diaganswer1}
&&
\Gamma^{(1)}_\infty=
\frac{1}{16\pi^2(d-4)}\ tr \int d^4 x\ \sqrt{-g} <
-\frac{(2l-1)(l^2-1)}{2(2l+1)}
\nabla_\mu \hat S^{\mu\nu} \nabla_\nu \hat S
+\frac{l^2}{2(2l+1)} \nabla_\mu \hat S\ \nabla^\mu \hat S
\nonumber\\
&&
+\frac{(2l-1)^2 l}{4(2l+1)} \nabla_\mu \hat S^\mu \nabla_\nu \hat S^\nu
+\frac{(2l-1)}{3} \nabla_\mu \hat S^\mu \hat S\ \hat S
-\frac{(2l-1)}{3} \nabla_\mu \hat S\ \hat S\ \hat S^\mu
+(l-1)\ \nabla_\mu \hat S\ \hat W^\mu
\nonumber\\
&&
- \frac{(2l-1)}{2} \nabla_\mu \hat S^\mu \hat W
+\hat S\ \hat N
+ \frac{1}{2} \hat W^2 + \frac{1}{4} \hat S^4
- \hat W\ \hat S^2-\hat M
+ \frac{(2l-1)(2l-3)(l-1)}{12}\ \nabla_\mu \hat S^{\mu\alpha\beta}
R_{\alpha\beta}
\nonumber\\
&&
- \frac{(2l-1)l}{12}\ \nabla_\mu \hat S^\mu R
- \frac{(2l-3)(l-1)}{6}\ \hat W^{\alpha\beta} R_{\alpha\beta}
+ \frac{l}{6} \hat W R
+ \frac{(2l-1)^2 (l+2)}{24(2l+1)}\ R_{\alpha\beta}
\hat S^\alpha \hat S^\beta
\nonumber\\
&&
- \frac{l}{6} \hat S^2 R
+ \frac{(2l-1)(l-1)}{6}\ R_{\alpha\beta}
\hat S^{\alpha\beta} \hat S
-\frac{(2l-1)(l-1)^2 (l+2)}{12(2l+1)}\ R_{\mu\nu\alpha\beta}
\hat S^{\mu\alpha} \hat S^{\nu\beta}
+ \frac{l}{12} F_{\mu\nu} F^{\mu\nu}
\nonumber\\
&&
+ \frac{l(2l-1)}{6}\nabla_\mu \hat S_\nu F^{\mu\nu}
+ \frac{(2l-1)^2 (l+1)}{4(2l+1)} F_{\mu\nu} \hat S^\mu \hat S^\nu
+ l\ (\frac{1}{120} R^2 + \frac{1}{60} R_{\mu\nu} R^{\mu\nu})>.
\end{eqnarray}

\narrowtext

It is more convenient sometimes to use another form of the operator:

\begin{eqnarray}\label{minimal2}
&&
\vphantom{\frac{1}{2}}
D_{i}{}^{j} = \delta_{i}{}^{j}
K_0 {}^{\mu_1\mu_2\ldots \mu_{2l}}
\nabla_{\mu_1} \nabla_{\mu_2}\ldots
\nabla_{\mu_{2l}}
\nonumber\\
&&
\vphantom{\frac{1}{2}}
+\ S^{\mu_1\mu_2\ldots \mu_{2l-1}}{}_{i}{}^{j}
\ \nabla_{\mu_1} \nabla_{\mu_2}\ldots
\nabla_{\mu_{2l-1}}
\nonumber\\
&&
\vphantom{\frac{1}{2}}
+\ W^{\mu_1\mu_2\ldots \mu_{2l-2}}{}_{i}{}^{j}
\ \nabla_{\mu_1} \nabla_{\mu_2}\ldots
\nabla_{\mu_{2l-2}}
\\
&&
\vphantom{\frac{1}{2}}
+\ N^{\mu_1\mu_2\ldots \mu_{2l-3}}{}_{i}{}^{j}
\ \nabla_{\mu_1} \nabla_{\mu_2}\ldots
\nabla_{\mu_{2l-3}} \nonumber\\
&&+
\vphantom{\frac{1}{2}}
\ M^{\mu_1\mu_2\ldots \mu_{2l-4}}{}_{i}{}^{j}
\ \nabla_{\mu_1} \nabla_{\mu_2}\ldots
\nabla_{\mu_{2l-4}} +
\ldots,\nonumber
\end{eqnarray}

\noindent
where $K_0$ was defined in (\ref{k0}).

By the help of the same method we found the following answer for
the divergent part of the one-loop effective action:

\widetext

\begin{eqnarray}\label{diaganswer2}
&&
\Gamma^{(1)}_\infty=
\frac{1}{16\pi^2(d-4)}\ tr \int d^4 x\ \sqrt{-g}
<
-\frac{(2l-1)(l^2-1)}{2(2l+1)}
\nabla_\mu \hat S^{\mu\nu} \nabla_\nu \hat S
+\frac{l^2}{2(2l+1)} \nabla_\mu \hat S\ \nabla^\mu \hat S
\nonumber\\
&&
+\frac{(2l-1)^2 l}{4(2l+1)} \nabla_\mu \hat S^\mu \nabla_\nu \hat S^\nu
+\frac{(2l-1)}{3} \nabla_\mu \hat S^\mu \hat S\ \hat S
-\frac{(2l-1)}{3} \nabla_\mu \hat S\ \hat S\ \hat S^\mu
+(l-1)\ \nabla_\mu \hat S\ \hat W^\mu
\nonumber\\
&&
- \frac{(2l-1)}{2} \nabla_\mu \hat S^\mu \hat W
+\hat S\ \hat N + \frac{1}{2} \hat W^2 + \frac{1}{4} \hat S^4
- \hat W\ \hat S^2-\hat M
+ \frac{(2l-1)(2l-3)(l-1)}{6(l+1)} \nabla_\mu
\hat S^{\mu\alpha\beta} R_{\alpha\beta}
\nonumber\\
&&
- \frac{l^2(2l-1)}{6(l+1)} \nabla_\mu \hat S^\mu R
+ \frac{l^2}{3(l+1)} \hat W\ R
- \frac{(2l-3)(l-1)}{3(l+1)} \hat W^{\mu\nu} R_{\mu\nu}
- \frac{(2l-1)^2 (l-4)}{24(2l+1)} R_{\mu\nu} \hat S^\mu \hat S^\nu
\nonumber\\
&&
- \frac{l^2}{2(2l+1)} \hat S^2 R
- \frac{(2l-1)(l-1)^2 (l+2)}{12(2l+1)} R_{\mu\nu\alpha\beta}
\hat S^{\mu\alpha} \hat S^{\nu\beta}
+ \frac{(2l-1)(l-1)(l+2)}{6(2l+1)} R_{\mu\nu}
\hat S^{\mu\nu} \hat S
\nonumber\\
&&
+ \frac{l^2(-7l^2+4l+12)}{540} R_{\mu\nu} R^{\mu\nu}
+ \frac{l^2(3l^2+4l+2)}{1080} R^2
+ \frac{l^3}{12} F^{\mu\nu} F_{\mu\nu}
+\frac{(l+1)(2l-1)^2}{4(2l+1)} \hat S^\mu \hat S^\nu F_{\mu\nu}
\nonumber\\
&&
+\frac{l^2 (2l-1)}{3(l+1)} F^{\mu\nu} \nabla_\mu \hat S_\nu>.
\end{eqnarray}

\narrowtext


\section{Nonminimal operator on the flat background}

Let us consider a theory with an arbitrary nonminimal operator
(\ref{nonminimal}) in the flat space ($g_{\mu\nu}= \eta_{\mu\nu}$,
$\omega_\mu{}_i{}^j=0$).

The divergent graphs are the same as in the case of a minimal operator
(see Fig.~\ref{flatdiagrams}), but because of the difference of the
propagators the calculations are also different.

As earlier, the calculation of logarithmically divergent graphs is
the simplest. For example,

\begin{eqnarray}
&&(2d)_\infty =
\frac{i}{2(2\pi)^4}\ tr \int d^dk\ (Sk)\ (Kk)^{-1} (N(k-p))
\nonumber\\
&&
\times
\left.
\vphantom{\frac{1}{2}}
(K(k-p))^{-1}
\right|_\infty
= \frac{1}{16\pi^2(d-4)}\ tr <\hat S\ \hat N>.
\end{eqnarray}

\noindent
(We should remind, that here $\hat W = (Kn)^{-1} (Wn) \ne (Wn)$ and so on,
see (\ref{notations})).

The other logarithmically divergent graphs can be calculated in the
same way. The result coincides with (\ref{ln}).

Now let us consider linearly divergent graphs.

\begin{eqnarray}
(2g) =
\frac{i}{2(2\pi)^4}\ tr &&\int d^dk\ (Sk)(Kk)^{-1}\nonumber\\
&&\times(W(k-p))(K(k-p))^{-1}.
\end{eqnarray}

We should expand this expression into series over external momentum
$p$ and retain logarithmically divergent terms (It is easy to see
that they are linear in $p$).

Using an identity

\begin{eqnarray}
1_i{}^j = (K(k+p))_i{}^m (K(k+p))^{-1}{}_m{}^j,
\end{eqnarray}

\noindent
we can easily obtain that

\begin{eqnarray}
&&
\vphantom{\frac{1}{2}}
(K(k+p))^{-1} = (Kk)^{-1} -\ L\ p_\mu (Kk)^{-1} (Kk)^\mu (Kk)^{-1}
\nonumber\\
&&+\ p_\mu p_\nu \left(
-\ \frac{1}{2} L(L-1)\ (Kk)^{-1} (Kn)^{\mu\nu} (Kk)^{-1}
\right.
\\
&&\left.
\vphantom{\frac{1}{2}}
+\ L^2 (Kk)^{-1} (Kn)^\mu (Kk)^{-1} (Kk)^\nu (Kk)^{-1} \right) +O(p^3).
\nonumber
\end{eqnarray}

\noindent
So, the divergent part of the diagram takes the following form

\begin{eqnarray}
\frac{1}{16\pi^2(d-4)}\ tr <L\ p_\mu &&\hat S\ \hat W\ \hat K^\mu
\nonumber\\
&&- (L-2)\ p_\mu \hat S\ \hat W^\mu >.
\end{eqnarray}

\noindent
Making a substitution $p_\mu \hat S \rightarrow -\ \partial_\mu \hat S $,
we found the ultimate expression in the coordinate space

\begin{eqnarray}
(2g)_\infty = \frac{1}{16\pi^2(d-4)}\ tr <&&
- L\ \partial_\mu \hat S\ \hat W
\ \hat K^\mu  \nonumber\\
&&+ (L-2)\ \partial_\mu \hat S\ \hat W^\mu >.
\end{eqnarray}

In a similar fashion we find that the divergent part of the graph
with three external $S$-lines

\widetext

\begin{eqnarray}
(2h) =
-\ \frac{i}{6(2\pi)^4}\ tr \int d^dk
\ (Sk)\ (Kk)^{-1}\ &&(S(k+q))
\times
\\
&&
\times
(K(k+q))^{-1}\ (S(k-p))\ (K(k-p))^{-1}\nonumber
\end{eqnarray}

\noindent
is

\begin{eqnarray}
(2h)_\infty = \frac{1}{16\pi^2(d-4)}&&\ tr <
\frac{1}{3} \left(
\vphantom{\frac{1}{2}}
(L-1)\ \partial_\mu \hat S^\mu  \hat S\
\hat S
-\right.\nonumber\\
&&\left.
-\ L\ \partial_\mu \hat S\  \hat K^\mu  \hat S\ \hat S\
\vphantom{\frac{1}{2}}
-\ (L-1)\ \partial_\mu \hat S\
\hat S\  \hat S^\mu
\vphantom{\frac{1}{2}}
+\ L\ \partial_\mu \hat S\  \hat S
\  \hat S\  \hat K^\mu
\right)>.
\end{eqnarray}

The divergent part of the diagram (2f)

\begin{eqnarray}
(2f)
= \frac{i}{4(2\pi)^4} tr \int d^dk\ (Sk)(Kk)^{-1}(S(k-p))(K(k-p))^{-1}
\end{eqnarray}

\noindent
should be calculated by series expansion in powers of $p$ to the
second order. It is easy to see, that

\begin{eqnarray}
(2f)_\infty = &&\frac{1}{16\pi^2(d-4)}\ tr <
-\ \frac{1}{2}\ \partial_\mu \hat S\ \partial_\nu \hat S
\left( -\ \frac{1}{2} L(L-1)\
\hat K^{\mu\nu}
\right.
+\nonumber\\
&&
\left.
\vphantom{\frac{1}{2}}
+\ L^2\ \hat K^\mu \hat K^\nu\right)
+\ \frac{1}{2} L(L-1)\ \partial_\mu \hat S\  \partial_\nu
\hat S^\nu  \hat K^\mu  -
\frac{1}{4} (L-1)(L-2)\ \partial_\mu \hat S\  \partial_\nu
\hat S^{\mu\nu}>.
\end{eqnarray}

Collecting contributions of all graphs we obtain the following
expression for the divergent part of the one-loop effective action

\begin{eqnarray}\label{nondiagflat}
&&
\left( \Gamma^{(1)}_\infty \right)^{flat} = \frac{1}{16\pi^2(d-4)}
tr \int d^4 x <
\frac{1}{4}\ \hat S^4
- \hat W\ \hat S^2
+\ \frac{1}{2}\ \hat W^2
+ \hat S\ \hat N
- \hat M
-\ L\ \partial_\mu \hat S\ \hat W\ \hat K ^\mu
\nonumber\\
&&
+(L-2)\ \partial_\mu \hat S\ \hat W ^\mu
\vphantom{\frac{1}{2}}
+\frac{1}{3} \left(
\vphantom{\frac{1}{2}}
(L-1)\ \partial_\mu \hat S^\mu \hat S\ \hat S
- L\ \partial_\mu \hat S\ \hat K^\mu \hat S\ \hat S
- (L-1)\ \partial_\mu \hat S\ \hat S\ \hat S^\mu
\right.
\nonumber\\
&&
\left.
\vphantom{\frac{1}{2}}
+ L\ \partial_\mu \hat S\ \hat S\ \hat S\ \hat K ^\mu
\right)
- \frac{1}{2}\ \partial_\mu \hat S
\ \partial_\nu \hat S
\left( - \frac{1}{2} L(L-1)\ \hat K ^{\mu\nu}
+ L^2 \hat K ^\mu \hat K ^\nu \right)
+ \frac{1}{2}L(L-1)\ \partial_\mu \hat S\ \partial_\nu
\hat S ^\nu \hat K ^\mu
\nonumber\\
&&
-\frac{1}{4}(L-1)(L-2)\ \partial_\mu \hat S\ \partial_\nu
\hat S ^{\mu\nu}>.
\end{eqnarray}

\narrowtext

This result can be generalized to the curved space-time. It will be
considered in our subsequent papers.


\section{Examples}

\subsection{The minimal second order operator}

If $l=1$ the operator (\ref{minimal1}) takes the form

\begin{equation}
D_{i}{}^{j} = \delta_{i}{}^{j} \Box + S^\mu{}_{i}{}^{j} \nabla_\mu +
W_{i}{}^{j}.
\end{equation}

It is easy to see that (\ref{diaganswer1}) gives then the following
expression

\widetext

\begin{eqnarray}
&&
\Gamma^{(1)}_\infty = \frac{1}{16\pi^2(d-4)}
\ tr \int d^4 x\ \sqrt{-g}
 <\frac{1}{6} \nabla_\mu \hat S\ \nabla^\mu \hat S +
+ \frac{1}{12} \nabla_\mu S^\mu \nabla_\nu S^\nu
- \frac{1}{2} \nabla_\mu S^\mu W
\nonumber\\
&&
+ \frac{1}{3} \nabla_\mu S^\mu \hat S\ \hat S
- \frac{1}{3} \nabla_\mu \hat S\ \hat S\ S^\mu
+ \frac{1}{2} W^2
+ \frac{1}{4} \hat S^4 - W \hat S^2
+ \frac{1}{120} R^2
+ \frac{1}{60} R_{\mu\nu} R^{\mu\nu}
+ \frac{1}{6} W R
\\
&&
- \frac{1}{12} \nabla_\mu S^\mu R
- \frac{1}{6} \hat S^2 R
+ \frac{1}{24} R_{\mu\nu} S^\mu S^\nu
+ \frac{1}{6} S^\mu S^\nu F_{\mu\nu}
+ \frac{1}{6} \nabla^\mu S^\nu F_{\mu\nu}
+ \frac{1}{12} F^{\mu\nu} F_{\mu\nu}
>. \nonumber
\end{eqnarray}

\narrowtext

\noindent
(Here $\hat W=W$,\ $\hat S^\mu=S^\mu$,\ $\hat S=S^\alpha n_\alpha$.)

After substituting according to (\ref{angle})

\begin{eqnarray}\label{middle}
&&<1> = 1,\qquad
<n_\mu n_\nu> = \frac{1}{4} g_{\mu\nu}, \\
&&<n_\mu n_\nu n_\alpha n_\beta> =
\frac{1}{24} (g_{\mu\nu} g_{\alpha\beta} +
g_{\mu\alpha} g_{\nu\beta} + g_{\mu\beta} g_{\nu\alpha}),\nonumber
\end{eqnarray}

\noindent
we obtain the well-known result \cite{thooft}:

\begin{eqnarray}\label{secondorder}
\Gamma^{(1)}_\infty = \frac{1}{16\pi^2(d-4)}
\ tr \int d^4 x\ \sqrt{-g}
\left( \frac{1}{12} Y_{\mu\nu} Y^{\mu\nu}
\right.
\nonumber\\
+ \frac{1}{2} X^2
+\frac{1}{60} R_{\mu\nu} R^{\mu\nu}
\left.
- \frac{1}{180} R^2 \right),
\end{eqnarray}

\noindent
where

\begin{eqnarray}
&&Y_{\mu\nu} = \frac{1}{2} \nabla_\mu S_\nu -
\frac{1}{2} \nabla_\mu S_\nu + \frac{1}{4} S_\mu S_\nu -
\frac{1}{4} S_\nu S_\mu + F_{\mu\nu},\nonumber\\
&&X = W - \frac{1}{2} \nabla_\mu S^\mu - \frac{1}{4} S_\mu S^\mu +
\frac{1}{6} R.
\end{eqnarray}


\subsection{The minimal forth order operator}

A minimal forth order operator $(l=2)$ has the form

\begin{eqnarray}\label{forthoper1}
D_{i}{}^{j} = \delta_{i}{}^{j} \Box^2 +
S^{\mu\nu\alpha}{}_{i}{}^{j}\ \nabla_\mu \nabla_\nu \nabla_\alpha +
W^{\mu\nu}{}_{i}{}^{j}\ \nabla_\mu \nabla_\nu
\nonumber\\
+ N^{\mu}{}_{i}{}^{j}\ \nabla_{\mu} +
M_{i}{}^{j}.
\end{eqnarray}

>From (\ref{diaganswer1}) we obtained the result, that coincided with
the one found in \cite{barv} by Barvinsky and Vilkovisky up to the total
derivatives. (We do not presented it here because it is too large)

In particular if $S^{\mu\nu\alpha} = 0$ the answer takes a rather
simple form

\begin{eqnarray}
&&
\Gamma^{(1)}_\infty =
\frac{1}{16\pi^2(d-4)}\ tr \int d^4 x\ \sqrt{-g}
\left(
\frac{1}{48} W^2
\right.
\nonumber\\
&&
+ \frac{1}{24} W^{\mu\nu} W_{\mu\nu}
- M +\frac{1}{60} R^2 + \frac{1}{30} R_{\mu\nu} R^{\mu\nu}
\\
&&
\left.
-\frac{1}{6} W^{\mu\nu} R_{\mu\nu} + \frac{1}{12} W R
+ \frac{1}{6} F_{\mu\nu} F^{\mu\nu}
\right).\nonumber
\end{eqnarray}


\subsection{The minimal sixth and eighth order operators}

Let us consider the most general sixth order operator without
fifth derivatives

\begin{eqnarray}
&&
D_i{}^j = \Box^3 \delta_i{}^j
+ W^{\mu\nu\alpha\beta}{}_i{}^j
\nabla_\mu \nabla_\nu \nabla_\alpha \nabla_\beta
+ N^{\mu\nu\alpha}{}_i{}^j \nabla_\mu \nabla_\nu \nabla_\alpha
\nonumber\\
&&
+ M^{\mu\nu}{}_i{}^j \nabla_\mu \nabla_\nu
\vphantom{\frac{1}{2}}
+ P^{\mu}{}_i{}^j \nabla_\mu + Q_i{}^j.
\end{eqnarray}

\noindent
(If $S \ne 0$ the result will be too cumbersome)

If $S=0$ the answer (\ref{diaganswer1}) gives

\widetext

\begin{eqnarray}\label{withoutS}
\Gamma^{(1)}_\infty=
\frac{1}{16\pi^2(d-4)}\ tr &&\int d^4 x\ \sqrt{-g}
< \frac{1}{2} \hat W^2 -\hat M
+ \frac{l}{12} F_{\mu\nu} F^{\mu\nu}
+ \frac{l}{6} \hat W R
-\\
&&
-\ \frac{1}{6}(2l-3)(l-1)\ \hat W^{\alpha\beta} R_{\alpha\beta}
+ \frac{l}{120} R^2 + \frac{l}{60} R_{\mu\nu} R^{\mu\nu}>.\nonumber
\end{eqnarray}

For the sixth order operator $l=3$, and, moreover, according to (\ref{wwa})
and (\ref{wwb})

\begin{eqnarray}
&&
<\hat M>\ = \frac{1}{4} M,
\qquad
<\hat W^{\mu\nu}>\ = \frac{1}{4} W^{\mu\nu},
\qquad
<\hat W>\ = \frac{1}{8} W,
\\
&&
<\hat W^2>\ =
\frac{1}{80} W^{\mu\nu\alpha\beta} W_{\mu\nu\alpha\beta}
+ \frac{3}{80} W^{\mu\nu} W_{\mu\nu}
+ \frac{3}{640} W^2,
\nonumber
\end{eqnarray}

\noindent
where $M\ \equiv M^{\mu\nu}{}_{\mu\nu}$ and so on.

So, for the sixth order operator we have

\begin{eqnarray}\label{six}
\Gamma^{(1)}_\infty=
&&\frac{1}{16\pi^2(d-4)}\ tr \int d^4 x\ \sqrt{-g}
\left(
\frac{1}{160} W^{\mu\nu\alpha\beta} W_{\mu\nu\alpha\beta}
+ \frac{3}{160} W^{\mu\nu} W_{\mu\nu}
\right.
+\\
&&
\left.
+ \frac{3}{1280} W^2
-\frac{1}{4} M
- \frac{1}{4} W^{\alpha\beta} R_{\alpha\beta}
+ \frac{1}{16} W R
+ \frac{1}{40} R^2 + \frac{1}{20} R_{\mu\nu} R^{\mu\nu}
+ \frac{1}{4} F_{\mu\nu} F^{\mu\nu}
\right).\nonumber
\end{eqnarray}

Eighth order operator

\begin{eqnarray}
&&
D_i{}^j = \Box^4 \delta_i{}^j
+ W^{\mu\nu\alpha\beta\gamma\delta}{}_i{}^j
\nabla_\mu \nabla_\nu \nabla_\alpha \nabla_\beta \nabla_\gamma \nabla_\delta
+ N^{\mu\nu\alpha\beta\gamma}{}_i{}^j
\nabla_\mu \nabla_\nu \nabla_\alpha \nabla_\beta \nabla_\gamma
+\\
&&
\vphantom{\frac{1}{2}}
+ M^{\mu\nu\alpha\beta}{}_i{}^j
\nabla_\mu \nabla_\nu \nabla_\alpha \nabla_\beta
+ P^{\mu\nu\alpha}{}_i{}^j \nabla_\mu \nabla_\nu \nabla_\alpha
+ Q^{\mu\nu}_i{}^j \nabla_\mu \nabla_\nu
+ T^{\mu}_i{}^j \nabla_\mu
+ V_i{}^j
\nonumber
\end{eqnarray}

\noindent
can be considered in a similar fashion:

In this case using rules formulated in the appendix \ref{diagangle},
in particular (\ref{wwa}) and (\ref{wwb}) we have

\begin{eqnarray}
&&
<\hat M>\ = \frac{1}{8} M,
\qquad
<\hat W^{\mu\nu}>\ = \frac{1}{8} W^{\mu\nu},
\qquad
<\hat W>\ = \frac{5}{64} W,
\\
&&
<\hat W^2>\ =
\frac{1}{14336}
\left(
16\ W^{\mu\nu\alpha\beta\gamma\delta}
W_{\mu\nu\alpha\beta\gamma\delta}
+ 120\ W^{\mu\nu\alpha\beta} W_{\mu\nu\alpha\beta}
\right.
+\nonumber\\
&&
\left.
\vphantom{\frac{1}{2}}
+ 90\ W^{\mu\nu} W_{\mu\nu}
+ 5\ W^2\right),
\nonumber
\end{eqnarray}

\noindent
and hence,

\begin{eqnarray}\label{eight}
&&
\Gamma^{(1)}_\infty=
\frac{1}{16\pi^2(d-4)}\ tr \int d^4 x\ \sqrt{-g}
\left(
\frac{1}{14336}
\left(
16\ W^{\mu\nu\alpha\beta\gamma\delta}
W_{\mu\nu\alpha\beta\gamma\delta}
\right.
+ 120\ W^{\mu\nu\alpha\beta} W_{\mu\nu\alpha\beta}
\right.
+\\
&&
\left.
+ 90\ W^{\mu\nu} W_{\mu\nu}
+ 5\ W^2\right)
-\frac{1}{8} M
- \frac{5}{16} W^{\alpha\beta} R_{\alpha\beta}
+ \frac{5}{96} W R
+ \frac{1}{30} R^2 + \frac{1}{15} R_{\mu\nu} R^{\mu\nu}
\left.
+ \frac{1}{3} F_{\mu\nu} F^{\mu\nu}
\right).\nonumber
\end{eqnarray}

\narrowtext


\subsection{Minimal operator as a special case of a nonminimal one}

A minimal operator can be considered as a particular case of a nonminimal
one, if

\begin{equation}
K^{\mu\nu\ldots\alpha}{}_i{}^j = \delta_i{}^j K_0^{\mu\nu\ldots\alpha}.
\end{equation}

In this case

\begin{eqnarray}
&&
\vphantom{\frac{1}{2}}
L = 2l, \qquad \qquad \quad(K_0 n) = 1,
\nonumber\\
&&\vphantom{\frac{1}{2}}
(K_0 n)^{-1} = 1, \qquad (K_0 n)^\mu = n^\mu,
\nonumber\\
&&
(K_0 n)^{\mu\nu} = \frac{1}{2l-1} \left(g^{\mu\nu} +(2l-2) n^\mu n^\nu
\right).
\nonumber
\end{eqnarray}

\noindent
Using the rules formulated in the appendix \ref{diagangle}. it is easy
to see that the result for the nonminimal operator (\ref{nondiagflat})
gives the answer (\ref{flat}) for the minimal operator (\ref{minimal1})
in the flat space.


\subsection{The nonminimal vector field operator}

In a lot of papers one can encounter the following nonminimal second
order operator

\begin{equation}\label{ndoper}
D_\alpha{}^\beta = \delta_{\alpha}{}^{\beta} \partial^\mu \partial_\mu
- \lambda\ \partial_\alpha \partial^\beta +
W_{\alpha}{}^\beta,
\end{equation}

\noindent
where

\begin{equation}
W_{\alpha\beta} = W_{\beta\alpha}.
\end{equation}

\noindent
It can be written in the form (\ref{nonminimal}) if

\begin{equation}
K^{\mu\nu}{}_\alpha{}^\beta = g^{\mu\nu} \delta_{\alpha}{}^{\beta} -
\frac{\lambda}{2}\ (g^{\mu\beta} \delta_\alpha{}^\nu +
g^{\nu\beta} \delta_\alpha{}^\mu),
\end{equation}

\noindent
and hence

\begin{eqnarray}
&&
(Kn)_\alpha{}^\beta =
\delta_\alpha{}^\beta - \lambda\ n_\alpha n^\beta,\nonumber\\
&&
(Kn)^{-1}{}_\alpha{}^\beta = \delta_\alpha{}^\beta + \gamma\ n_\alpha
n^\beta, \quad \mbox{where} \quad
\gamma \equiv \frac{\lambda}{1-\lambda}.
\end{eqnarray}

Substituting it to the equation (\ref{nondiagflat}) and
using (\ref{middle}) we found that

\begin{eqnarray}
&&
\Gamma^{(1)}_\infty =
\frac{1}{16\pi^2(d-4)}\ tr \int d^4 x\ \sqrt{-g}
\ \left(
\frac{1}{48} \gamma^2 W^2\right.\nonumber\\
&&+\left.
\left(\frac{1}{24} \gamma^2 + \frac{1}{4} \gamma +\frac{1}{2}\right)
W_{\mu\nu} W^{\mu\nu}\right),
\end{eqnarray}

\noindent
where $ W \equiv W_\alpha{}^\alpha$.

This expression is also in agreement with \cite{barv},
\cite{gusynin1} and \cite{fradkin}.


\section{Conclusion.}

In this paper we presented a new method for making one-loop calculations in
the nonminimal gauges \cite{kallosh}, theories regularized by higher
derivatives \cite{slavnov1} and field theory models with higher spins.

>From the mathematical point of view we found $a_2$ coefficient
of the heat kernel expansion of an arbitrary differential
operator without any restriction to its form and order.
As a tool we choosed t'Hooft-Veltman diagramic approach with some
modifications, in particular in the technick of calculating Feynman
integrals.

Unfortunately, in the general case we can not use the classical gauge
invariance that reliefs calculations considerably for the minimal second
order operator \cite{thooft2}. However we manage to obtain master formulas
for the one-loop renormalization counterterms in theories with an
arbitrary minimal operator in the curved space and an arbitrary
nonminimal operator in the flat space. The consideration of some particular
cases showed the agreement of our results with algorithms obtained earlier.

The result for a nonminimal operator can be generalized to the
curved space. This will be reported in a separate paper. Of course, it
is also possible to extend the method to multi-loop orders.

On the other hand, our approach can be considered as a step towards the
total automatization of the one-loop divergences calculation. We also
tries to create the appropriate software. For example, making calculation
we often used our tensor package for the REDUCE system \cite{tensor}.
So, the most cumbersome operations in calculating one-loop Feynman diagrams
now can be made by computers.


\appendix

\section{
Calculation of the divergent part of Feynman integrals.
}\label{integ}

Let us consider an integral

\begin{equation}\label{int}
\int d^{d}k\ f(k,p_{1},p_{2}\ldots p_{n}),
\end{equation}

\noindent
where $d$ is the space-time dimension.

Expanding $f$ in powers of $1/k$ we have

\begin{eqnarray}
\int d^{d}k\ f(k,p_{1}\ldots p_{n}) = \int d^{d}k
\left(\frac{1}{k^{(S)}} f_{S}(p_{1}\ldots p_{n})\right.\nonumber\\
\left.
+ \frac{1}{k^{(S+1)}} f_{S+1}(p_{1}\ldots p_{n}) + \ldots\right).
\end{eqnarray}

Here $1/k^{(S)}$ is an arbitrary expression built by $k_{\mu}$
of order $-S$, For example, by $1/k^{(2)}$ we denote

\[ \frac{1}{k^2}, \frac{k_{\mu}}{k^3}, \frac{k_{\mu}
k_{\nu}}{k^4}, \mbox{\ and so on}\]

\noindent
where $k = \sqrt{k^{2}}$.

The integrals are divergent if $s \le 4$. In this case we first perform
the angle integration.

It is easy to see that

\[\int d^{d}k\ \frac{1}{k^{L+S}}
k_{\mu_{1}} k_{\mu_{2}} \ldots  k_{\mu_{L}} \]

\noindent
is equal to 0, if $L = 2l+1$ and

\begin{eqnarray}\label{intrc}
C \int d^{d}k\ \frac{1}{k^{S}}
&&\left(g_{\mu_{1}\mu_{2}} g_{\mu_{3}\mu_{4}}\ldots g_{\mu_{L-1}\mu_{L}}
\right.
\nonumber\\
&&\left.
+g_{\mu_{1}\mu_{3}} g_{\mu_{2}\mu_{4}}\ldots g_{\mu_{L-1}\mu_{L}} + \ldots
\right),
\end{eqnarray}

\noindent
if $L = 2l$

Here the sum is over all possible index replacements. It contains
$(2l-1)!! = (2l-1) (2l-3) \ldots 1$ terms.
($(-1)!! = 1$) Contracting (\ref{intrc}) with
$g_{\mu_{1}\mu_{2}} g_{\mu_{3}\mu_{4}}\ldots g_{\mu_{L-1}\mu_{L}}$
we found a constant $C$

\begin{equation}
C = \frac{1}{2^{l} (l+1)!}.
\end{equation}

Therefore, introducing a notation

\begin{eqnarray}\label{angleprl}
< n_{\mu_{1}} n_{\mu_{2}} \ldots n_{\mu_{2l}}&&>\ \equiv
\frac{1}{2^{l} (l+1)!}
\nonumber\\
\times
\vphantom{\frac{1}{2}}
\left(g_{\mu_{1}\mu_{2}} g_{\mu_{3}\mu_{4}}
\right. \ldots &&g_{\mu_{2l-1}\mu_{2l}}
\nonumber\\
\vphantom{\frac{1}{2}}
+
&&\left.
g_{\mu_{1}\mu_{3}} g_{\mu_{2}\mu_{4}}\ldots g_{\mu_{2l-1}\mu_{2l}} + \ldots
\right),
\end{eqnarray}

\noindent
we obtain

\begin{eqnarray}\label{angle2}
\int d^{d}k\ \frac{1}{k^{2l+S}}
&&k_{\mu_{1}} k_{\mu_{2}} \ldots  k_{\mu_{2l}} \nonumber\\
&&
= \ < n_{\mu_{1}} n_{\mu_{2}} \ldots  n_{\mu_{2l}} >
\int d^{d}k\ \frac{1}{k^{S}}.
\end{eqnarray}

Using

\begin{equation}
\int d^{d}k\ \frac{1}{(k^{2}+L)^\alpha} =
\pi^{d/2} e^{i \pi/2 - i \pi d/2}
\frac{\Gamma(\alpha - d/2)}{\Gamma(\alpha)}
L^{d/2 - \alpha},
\end{equation}

\noindent
we conclude that nonzero result can be obtained if only
$d/2 -\alpha =0$ (or $s = 2 \alpha = 4$). Then the divergent part will
be the following:

\begin{equation}
\left(\int d^{d}k\ \frac{1}{k^{4}}\right)_{\infty} =
\left(\lim_{L\to 0} \int d^{d}k\ \frac{1}{(k^{2}+L)^2}\right)_{\infty}
= - \frac{2 i \pi^{2}}{d-4},
\end{equation}

\noindent
and

\begin{eqnarray}\label{angle3}
\left(\int d^{d}k\ \frac{1}{k^{2l+S}}
k_{\mu_{1}} k_{\mu_{2}} \ldots  k_{\mu_{2l}}\right)_\infty &&\nonumber\\
= -\ \frac{2 i\pi^2}{d-4}\ <n_{\mu_{1}} n_{\mu_{2}}&& \ldots n_{\mu_{2l}}>.
\end{eqnarray}

Therefore, in order to calculate the divergent part of the integral
(\ref{int}) we should expand $f$ in powers of $1/k$, retain
logarithmically divergent terms and perform the remaining integration by
(\ref{angle3}).


\section{The integration over angles for a minimal operator}
\label{diagangle}

Let us start with (\ref{angleprl}):

\begin{eqnarray}\label{angle1}
< n_{\mu_{1}} n_{\mu_{2}} \ldots n_{\mu_{2m}}&&>\ \equiv
\frac{1}{2^{m} (m+1)!}
\nonumber\\
\times
\vphantom{\frac{1}{2}}
\left(g_{\mu_{1}\mu_{2}} g_{\mu_{3}\mu_{4}}
\right. \ldots &&g_{\mu_{2m-1}\mu_{2m}}
\nonumber\\
+&&
\vphantom{\frac{1}{2}}
\left.
g_{\mu_{1}\mu_{3}} g_{\mu_{2}\mu_{4}}\ldots g_{\mu_{2m-1}\mu_{2m}} + \ldots
\right).
\end{eqnarray}

\noindent
It can be interpreted in the following way:
In order to obtain the result of the angle integration we should
make pairs of $n_\alpha$ by all possible ways and add a numerical constant.
Each pair  of $n_\alpha$ and $n_\beta$ should be substituted by
$g_{\alpha\beta}$.

The sum contains $(2m-1)!!$ terms. Hence, if we contract (\ref{angle1})
with a totally symmetric tensor
$A^{\mu_1\mu_2\ldots\mu_{2m}} \equiv A_{(2m)}$
(here the bottom index points the tensor rank) the result will be

\begin{equation}\label{rule1}
<(A_{(2m)}n)>\ =  \frac{(2m-1)!!}{2^m(m+1)!}\ A,
\end{equation}

\noindent
where $A\ \equiv A^{\mu_1\ldots\mu_m}{}_{\mu_1\ldots\mu_m}$.

Similarly one can find that for a symmetric tensor $A_{(2m-1)}$ with
$2m-1$ indexes the following equation takes place:

\begin{eqnarray}\label{rule2}
<n_\alpha (A_{(2m-1)}n)>\ =  \frac{(2m-1)!!}{2^m(m+1)!}\ A_\alpha
\nonumber\\
= \frac{2m-1}{2(m+1)}\ <(A_{(2m-1)}n)_\alpha>,
\end{eqnarray}

\noindent
where $A_\alpha\ \equiv
A^{\mu_1\ldots\mu_{m-1}}{}_{\mu_1\ldots\mu_{m-1}\alpha}$.

In the more general case we will use the following consequence of
(\ref{angle}):

\widetext

\begin{eqnarray}
&&<n_{\mu_1} n_{\mu_2} \ldots n_{\mu_{2m}}>\ = \frac{1}{2(m+1)}
\left( \vphantom{\sqrt{1}}g_{\mu_1\mu_2}
<n_{\mu_3} n_{\mu_4} \ldots n_{\mu_{2m}}> \right. \\
&&+ \left. g_{\mu_1\mu_3}
<n_{\mu_2} n_{\mu_4} \ldots n_{\mu_{2m}}> + \ldots + g_{\mu_1\mu_{2m}}
<n_{\mu_2} n_{\mu_3} \ldots n_{\mu_{2m-1}}>\right).\nonumber
\end{eqnarray}

Making contraction with 2 symmetric tensors we find that

\begin{eqnarray}\label{eqa}
<n_\alpha (A_{(2m)}n) &&(B_{(2p-1)}n)>\ =
\frac{1}{2(m+p+1)}
\\
&&
\times
\left(
2m <(A_{(2m)}n){}_\alpha (B_{(2p-1)}n)>
+\ (2p-1) <(A_{(2m)}n)
(B_{(2p-1)}n){}_\alpha>\right).
\vphantom{\frac{1}{2}}
\nonumber
\end{eqnarray}

\noindent
This equation can be easily generalized to a greater number of
symmetric tensors.

For a minimal operator we can explicitly make angle integration, although
a result will be very cumbersome. Nevertheless, it is very convenient
for computer calculations, because a number of operations to do becomes very
small.

First we consider

\begin{equation}\label{wwa}
<(A_{(2m)}n) (B_{(2p)}n)>\ = \sum_{k=0}^{m}
A^{\mu_1 \ldots \mu_{2k}} B_{\mu_1 \ldots \mu_{2k}} T_{mp}^k
\end{equation}

\noindent
and assume that $m \le p$. $T_{mp}^k$ is a numeric constant that should be
calculated. It is easy to see, that

\begin{eqnarray}\label{wwb}
&&
T_{mp}^k =
\frac{1}{2^{m+p}(m+p+1)!}
\ C^{2k}_{2m}\ (2m-2k-1)!!\ (2p-2k-1)!!\ \frac{(2p)!}{(2p-2k)!}.
\end{eqnarray}

\noindent
Really,
$1/2^{m+p}(m+p+1)!$ is a normalization constant,
$C^{2k}_{2m}$ gives a number of ways to choose $2k$ $n_\alpha$ from
$(A_{(2m)}n)$,
$(2p)!/(2p-2k)!$ is a number of their possible combinations with
$n_\beta$ from $(B_{(2p)}n)$ and
$(2m-2k-1)!!\ (2p-2k-1)!!$ is a number of $n_\mu$ pairs inside
$(A_{(2m)}n)$ and $(B_{(2p)}n)$.

By similar arguments

\begin{equation}\label{eqb}
<(A_{(2m+1)}n) (B_{(2p+1)}n)>\ = \sum_{k=0}^{m} A^{\mu_1 \ldots \mu_{2k+1}}
B_{\mu_1 \ldots \mu_{2k+1}} T_{mp}^k,
\end{equation}

\noindent
where

\begin{eqnarray}
&&
T_{mp}^k = \frac{1}{2^{m+p+1}(m+p+2)!}
\ C^{2k+1}_{2m+1} (2m-2k-1)!!\ (2p-2k-1)!!
\frac{(2p+1)!}{(2p-2k)!}
\end{eqnarray}

\noindent
and $m \le p$.

(\ref{eqa}) and (\ref{eqb}) can be written as

\begin{equation}
<(A_{(m)}n) (B_{(p)}n)>\ = \sum_{k=0}^{m} A^{\mu_1 \ldots \mu_{k}}
B_{\mu_1 \ldots \mu_{k}} T_{mp}^k,
\end{equation}

\noindent
where $m \le p$ and

\begin{eqnarray}
T_{mp}^k =
\frac{1}{{\displaystyle 2^{(m+p)/2}}
{\textstyle ((m+p)/2+1)!}}
C^{k}_{m}\ (m-k-1)!!\ (p-k-1)!!\ \frac{p!}{(p-k)!},
\end{eqnarray}

\noindent
if $m-k$ and $p-k$ are both even and $0$, otherwise.

This equation can be generalized to an arbitrary number of
tensors. Here we consider cases that encounter in (\ref{diaganswer1})
and (\ref{diaganswer2}). Some terms there have a form

\begin{eqnarray}
&&<(A_{(2l-2)}n) (B_{(2l-1)}n) (C_{(2l-1)}n)> \\
&&=\sum_{m_1=0}^{2l-2} \sum_{m_2=0}^{2l-2} \sum_{m_3=0}^{2l-1}
A^{\mu_1\ldots\mu_{m_1}\nu_1\ldots\nu_{m_2}}
B^{\mu_1\ldots\mu_{m_1}\alpha_1\ldots\alpha_{m_3}}
C^{\nu_1\ldots\nu_{m_2}\alpha_1\ldots\alpha_{m_3}}
T_l^{m_1 m_2 m_3}.\nonumber
\end{eqnarray}

\noindent
By the help of the same method we can find that

\begin{eqnarray}
&&T_l^{m_1 m_2 m_3}=
\frac{1}{{\displaystyle 2^{3l-2}}
{\textstyle (3l-1)!}}
\vphantom{\frac{1}{2}}
\ C^{m_1}_{2l-2}\ C^{m_2}_{2l-2-m_1}\ C^{m_3}_{2l-1-M}
(2l-2-m_1-m_3)!!
\\
&&\times
\vphantom{\frac{1}{2}}
(2l-2-m_2-m_3)!!
(2l-3-m_1-m_2)!!
\frac{(2l-1)!}{(2l-1-m_1)!}
\ \frac{(2l-1)!}{(2l-1-m_2)!}
\frac{(2l-1-m)!}{(2l-1-m-m_3)!},
\nonumber
\end{eqnarray}

\noindent
if $m_1+m_2$ is even and $m_1+m_3$, $m_2+m_3$ is odd,
otherwise $T_l^{m_1 m_2 m_3}= 0$
($M \equiv max(m_1,m_2)$ É $m \equiv min(m_1,m_2)$).

\noindent
A term $<(S_{(2l-1)}n)^4>$ can be written as

\begin{eqnarray}
&&\vphantom{\frac{1}{2}}<(S_{(2l-1)}n)^4> =
\sum_{k_1=0}^{2l-1}\ \sum_{k_2=0}^{2l-1-k_1}
\ \sum_{k_3=0}^{2l-1-k_1-k_2}\ \sum_{k_3=0}^{2l-1-M_1}
\ \sum_{k_3=0}^{2l-1-M_2}\ \sum_{k_3=0}^{2l-1-M_3}
\\
&&
\vphantom{\frac{1}{2}}
\ S^{\mu_1\ldots\mu_{k_1}\nu_1\ldots\nu_{k_2}\sigma_1\ldots\sigma_{k_3}}
\vphantom{\frac{1}{2}}
S^{\mu_1\ldots\mu_{k_1}\alpha_1\ldots\alpha_{k_4}\beta_1\ldots\beta_{k_5}}
 S^{\nu_1\ldots\nu_{k_2}\alpha_1\ldots\alpha_{k_4}
\gamma_1\ldots\gamma_{k_6}}
\nonumber\\
&&
\vphantom{\frac{1}{2}}
\times
\ S^{\sigma_1\ldots\sigma_{k_3}\beta_1\ldots\beta_{k_5}
\gamma_1\ldots\gamma_{k_6}}
\ T_l^{k_1 k_2 k_3 k_4 k_5 k_6},\nonumber
\end{eqnarray}

\noindent
where

\begin{eqnarray}
&&
T_l^{k_1 k_2 k_3 k_4 k_5 k_6}=
\frac{1}{4^{2l-1} (4l-1)!}
\ (2l-2-k_1-k_2-k_3)!!
\\
&&
\times
\vphantom{\frac{1}{2}}
(2l-2-k_1-k_4-k_5)!!
\ (2l-2-k_2-k_4-k_6)!!
\ (2l-2-k_3-k_5-k_6)!!
\nonumber\\
&&
\times
\ C^{k_1}_{2l-1}
\frac{(2l-1)!}{(2l-1-k_1)!}
C^{k_2}_{2l-1-k_1}
\frac{(2l-1)!}{(2l-1-k_2)!}
\ C^{k_3}_{2l-1-k_1-k_2}
\frac{(2l-1)!}{(2l-1-k_3)!}
\nonumber\\
&&
\times
C^{k_4}_{2l-1-M_1}
\frac{(2l-1-m_1)!}{(2l-1-m_1-k_4)!}
C^{k_5}_{2l-1-M_2}
\frac{(2l-1-m_2)!}{(2l-1-m_2-k_5)!}
\ C^{k_6}_{2l-1-M_3}
\frac{(2l-1-m_3)!}{(2l-1-m_3-k_5)!},
\nonumber
\end{eqnarray}

\noindent
if $k_1+k_2+k_3$,\ $k_1+k_4+k_5$,\ $k_2+k_4+k_6$,\ $k_3+k_5+k_6$
are odd. Here

\begin{equation}
\begin{array}{lll}
M_1\ \equiv max(k_1,k_2),& m_1\ \equiv min(k_1,k_2),\\
M_2\ \equiv max(k_1+k_4,k_3),& m_2\ \equiv min(k_1+k_4,k_3),\\
M_3\ \equiv max(k_2+k_4,k_3+k_5),\qquad& m_3\ \equiv min(k_2+k_4,k_3+k_5).
\end{array}
\end{equation}

Otherwise
\[T_l^{k_1 k_2 k_3 k_4 k_5 k_6}=0.\]

We should note that rules formulated here are applicable only for a
minimal operator. The matter is that for a nonminimal operator one
can not present in general $(Kn)^{-1}$ as a contraction of the form
$(Tn)$, where $T$ is a totally symmetric tensor.

\narrowtext


\pagebreak

\widetext

\begin{figure}

\unitlength=1pt

\begin{picture}(360,75)(50,0)
\thicklines

\put(110,30){\circle{29.9}}
\put(120.5,19.5){\circle*{4.2}}
\put(120.5,40.5){\circle*{4.2}}
\put(99.5,19.5){\circle*{4.2}}
\put(99.5,40.5){\circle*{4.2}}
\put(122.5,40.5)
{\thicklines
\multiput(0,0)(10.3,2.2){5}%
{\put(0,0){\oval(5,5)[b]}}
\multiput(5.3,0.3)(10.3,2.2){4}%
{\put(0,0){\oval(5,5)[t]}}
\multiput(1.6,0.1)(10.3,2.2){4}%
{.}}
\put(55.4,49.5)
{\thicklines
\multiput(0,0)(10.3,-2.2){5}%
{\put(0,0){\oval(5,5)[t]}}
\multiput(5.3,-0.3)(10.3,-2.2){4}%
{\put(0,0){\oval(5,5)[b]}}
\multiput(1.8,-0.35)(10.3,-2.2){4}%
{.}}
\put(122.5,19.5)
{\thicklines
\multiput(0,0)(10.3,-2.2){5}%
{\put(0,0){\oval(5,5)[t]}}
\multiput(5.3,-0.3)(10.3,-2.2){4}%
{\put(0,0){\oval(5,5)[b]}}
\multiput(1.8,-0.35)(10.3,-2.2){4}%
{.}}
\put(55.4,10.5)
{\thicklines
\multiput(0,0)(10.3,2.2){5}%
{\put(0,0){\oval(5,5)[b]}}
\multiput(5.3,0.3)(10.3,2.2){4}%
{\put(0,0){\oval(5,5)[t]}}
\multiput(1.6,0.1)(10.3,2.2){4}%
{.}}
\put(55.4,16.5){S}
\put(161,16.5){S}
\put(55.4,55.5){S}
\put(161,55.5){S}
\put(215,27){(2a)}

\put(370,30){\circle{29.9}}
\put(355,30){\circle*{4.2}}
\put(380.5,19.5){\circle*{4.2}}
\put(380.5,40.5){\circle*{4.2}}
\put(313,31.5){\line (1,0) {42}}
\put(313,28.5){\line (1,0) {42}}
\put(383,40.5)
{\thicklines
\multiput(0,0)(10.3,2.2){5}%
{\put(0,0){\oval(5,5)[b]}}
\multiput(5.3,0.3)(10.3,2.2){4}%
{\put(0,0){\oval(5,5)[t]}}
\multiput(1.6,0.1)(10.3,2.2){4}%
{.}}
\put(383,19.5)
{\thicklines
\multiput(0,0)(10.3,-2.2){5}%
{\put(0,0){\oval(5,5)[t]}}
\multiput(5.3,-0.3)(10.3,-2.2){4}%
{\put(0,0){\oval(5,5)[b]}}
\multiput(1.8,-0.35)(10.3,-2.2){4}%
{.}}
\put(313,39){W}
\put(421,55.5){S}
\put(421,16.5){S}
\put(480,27){(2b)}

\end{picture}

\begin{picture}(360,75)(50,0)
\thicklines

\put(110,30){\circle{29.9}}
\put(95,30){\circle*{4.2}}
\put(125,30){\circle*{4.2}}
\put(53,31.5){\line (1,0) {42}}
\put(53,28.5){\line (1,0) {42}}
\put(125,31.5){\line (1,0) {42}}
\put(125,28.5){\line (1,0) {42}}
\put(53,36){W}
\put(161,36){W}
\put(215,27){(2c)}

\put(370,30){\circle{29.9}}
\put(355,30){\circle*{4.2}}
\put(385,30){\circle*{4.2}}
\put(382,30){
\put(3,2.1){\line (1,0) {42}}
\put(3,0){\line (1,0) {42}}
\put(3,-2.1){\line (1,0) {42}}}
\put(309,30)
{\thicklines
\multiput(0,0)(9.2,0){5}%
{\put(0,-0.05){\oval(4,4)[t]}
\put(4.6,0.05){\oval(4,4)[b]}}}
\put(309,39){S}
\put(421,39){N}
\put(480,27){(2d)}

\end{picture}

\begin{picture}(360,75)(50,0)
\thicklines

\put(110,30){\circle{29.9}}
\put(95,30){\circle*{4.2}}
\put(50,30){
\put(3,1){\line (1,0) {42}}
\put(3,-1){\line (1,0) {42}}
\put(3,3){\line (1,0) {42}}
\put(3,-3){\line (1,0) {42}}
\put(3,9){M}
}
\put(215,27){(2e)}

\put(370,30){\circle{29.9}}
\put(355,30){\circle*{4.2}}
\put(385,30){\circle*{4.2}}
\put(309,30)
{\thicklines
\multiput(0,0)(9.2,0){5}%
{\put(0,-0.05){\oval(4,4)[t]}
\put(4.6,0.05){\oval(4,4)[b]}}}
\put(386,30)
{\thicklines
\multiput(0,0)(9.2,0){5}%
{\put(0,-0.05){\oval(4,4)[t]}
\put(4.6,0.05){\oval(4,4)[b]}}}
\put(310,39){S(-p)}
\put(409,39){S(p)}
\put(480,27){(2f)}
\put(370,45){\vector(1,0){0.1}}
\put(370,15){\vector(-1,0){0.1}}
\put(367,51){k}
\put(361,3){k-p}

\end{picture}

\begin{picture}(360,95)(50,0)
\thicklines

\put(110,50){\circle{29.9}}
\put(95,50){\circle*{4.2}}
\put(125,50){\circle*{4.2}}
\put(125,51.5){\line (1,0) {42}}
\put(125,48.5){\line (1,0) {42}}
\put(48.8,50)
{\thicklines
\multiput(0,0)(9.2,0){5}%
{\put(0,-0.05){\oval(4,4)[t]}
\put(4.6,0.05){\oval(4,4)[b]}}}
\put(48.8,59){S(-p)}
\put(143,59){W(p)}
\put(215,47){(2g)}
\put(110,65){\vector(1,0){3}}
\put(110,35){\vector(-1,0){3}}
\put(107,71){k}
\put(101,23){k-p}

\put(370,50){\circle{29.9}}
\put(355,50){\circle*{4.2}}
\put(380.5,39.5){\circle*{4.2}}
\put(380.5,60.5){\circle*{4.2}}
\put(309,50)
{\thicklines
\multiput(0,0)(9.2,0){5}%
{\put(0,-0.05){\oval(4,4)[t]}
\put(4.6,0.05){\oval(4,4)[b]}}}
\put(383,60.5)
{\thicklines
\multiput(0,0)(10.3,2.2){5}%
{\put(0,0){\oval(5,5)[b]}}
\multiput(5.3,0.3)(10.3,2.2){4}%
{\put(0,0){\oval(5,5)[t]}}
\multiput(1.6,0.1)(10.3,2.2){4}%
{.}}
\put(383,39.5)
{\thicklines
\multiput(0,0)(10.3,-2.2){5}%
{\put(0,0){\oval(5,5)[t]}}
\multiput(5.3,-0.3)(10.3,-2.2){4}%
{\put(0,0){\oval(5,5)[b]}}
\multiput(1.8,-0.35)(10.3,-2.2){4}%
{.}}
\put(480,47){(2h)}
\put(309,59){S(-p)}
\put(412,78.5){S(-q)}
\put(403,14){S(p+q)}
\put(385,50){\vector(0,-1){3}}
\put(367,64.7){\vector(4,1){3}}
\put(367,35.6){\vector(-4,1){3}}
\put(364,71){{\small k}}
\put(358,23){{\small k-p}}
\put(391,47){{\small k+q}}

\end{picture}

\caption{Divergent graphs in the flat space. Diagrams (2a) - (2e)
are logarithmically divergent, (2g) and (2h) are linearly divergent
and (2f) is a quadratically divergent graph.
}\label{flatdiagrams}
\end{figure}

\pagebreak

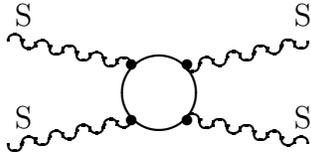
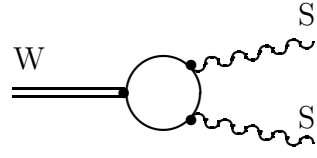
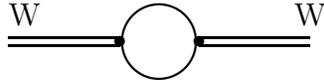
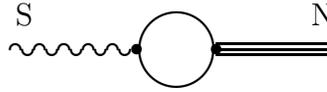
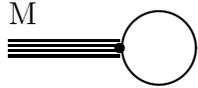
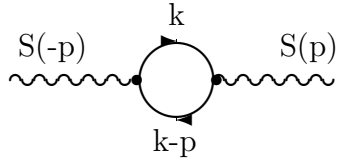
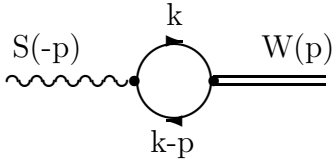
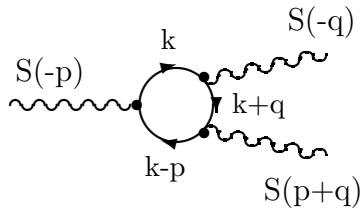
\end{document}